\documentclass[11pt]{article}
\usepackage{amstext,amssymb,amsmath,amsbsy}

\usepackage{hyperref}
\usepackage{amscd} 
\usepackage{amsfonts}
\usepackage{indentfirst}
\usepackage{verbatim}
\usepackage{amsmath}
\usepackage{amsthm}
\usepackage{enumerate}
\usepackage{graphicx}
\usepackage{subfig}
\usepackage{color} 
\usepackage[OT1]{fontenc}
\usepackage[latin1]{inputenc}
\usepackage[english]{babel}
\usepackage{amssymb}
\newtheorem{theorem}{Theorem}
\newtheorem{lemma}{Lemma}

\newtheorem{proposition}{Proposition}
\newtheorem{corollary}{Corollary}

\makeatletter
\def\th@newremark{\th@remark\thm@headfont{\bfseries}}
\makeatletter
\theoremstyle{newremark}
\newtheorem{remark}{Remark}
\newtheorem{notation}{Notation}

\newtheorem{definition}{Definition}
\numberwithin{equation}{section}
\numberwithin{lemma}{section}
\numberwithin{notation}{section}
\numberwithin{corollary}{section}
\numberwithin{corollary}{section}
\numberwithin{example}{section}
\numberwithin{definition}{section}
\numberwithin{remark}{section}

\newcommand{\proofend}{\hfill $\Box$ }

\newcommand{\dsp}{\displaystyle}

\newcommand{\supp}{\operatorname{supp}}

\newcommand{\dive}{\operatorname{div}}

\newcommand{\bH}{{\bf H}}
\newcommand{\hR}{ \hat R}

\newcommand{\eps}{\varepsilon}

\newcommand{\loc}{_{loc}}

\newcommand{\mN}{\mathbb{N}}

\newcommand{\mR}{\mathbb{R}}

\newcommand{\mc}{\mathrm{c}}

\newcommand{\cA}{{\hat A}}
\newcommand{\cU}{{\hat u}}
\newcommand{\cS}{\hat \Sigma}

\title{Cloaking an arbitrary object via anomalous localized resonance: the cloak is independent of the object}

\author{Hoai-Minh Nguyen \footnote{EPFL SB MATHAA CAMA, Station 8,  CH-1015 Lausanne, hoai-minh.nguyen@epfl.ch}}

\begin{document}

\maketitle

\begin{abstract}

In this paper, we present various schemes of cloaking an arbitrary objects via anomalous localized resonance and provide their analysis in two and three dimensions. This is a way to cloak an object using negative index materials in which the cloaking device is independent of the object. As a result, we show that in two dimensional quasistatic regime  an annular plasmonic structure  $-I$ cloaks small but finite size objects near by.  We also discuss its connections with superlensing and cloaking using complementary media. In particular, we confirm the possibility that a lens can act like a cloak and conversely. This possibility was raised about a decade ago in the literature.

\medskip 

\noindent{\bf Key words:} cloaking, superlensing, complementary media, localized resonance, three spheres inequality, conformal maps. 

\end{abstract}

\tableofcontents

\section{Introduction}

Negative index materials (NIMs) were  investigated theoretically by Veselago in \cite{Veselago}  and the  existence of such materials was confirmed by Shelby, 
Smith, and Schultz in \cite{ShelbySmithSchultz}. The study of NIMs has attracted a lot attention in the scientific community thanks to their many potential applications. Mathematically, the study of NIMs faces two difficulties.  First, the equations modeling NIMs have sign changing coefficients; hence the ellipticity and the compactness are lost in general. Secondly, the localized resonance, i.e. the fields blow up in some regions and remain bounded in some others as the loss goes to 0, might appear.


\medskip 
Three known applications of NIMs are superlensing and cloaking using complementary media and cloaking a source via anomalous localized resonance (ALR). Superlensing using complementary media was suggested by Veselago in \cite{Veselago} for a slab lens (a slab of index $-1$) using the ray theory.  Later, cylindrical lenses in the two dimensional  quasistatic regime, the Veselago slab lens  and cylindrical lenses in the finite frequency regime, and   spherical lenses in the finite frequency regime were studied by Nicorovici, McPhedran, and Milton  in \cite{NicoroviciMcPhedranMilton94},  Pendry in \cite{PendryNegative, PendryCylindricalLenses}, and Ramakrishna and Pendry  in \cite{PendryRamakrishna} respectively for  dipoles sources. Superlensing using complementary media for arbitrary objects in the acoustic and electromagnetic settings  was established by  Nguyen  in  \cite{Ng-Superlensing, Ng-Superlensing-Maxwell} for related schemes. Cloaking using  complementary media was suggested and investigated numerically by Lai et al. in \cite{LaiChenZhangChanComplementary}. This  was established rigorously for related schemes by Nguyen in \cite{Ng-Negative-Cloaking} for  the  quasi-static regime and later extended by Nguyen and (H. L.) Nguyen in \cite{MinhLoc2} for the finite frequency regime.  Cloaking a source via ALR was discovered by Milton and 
 Nicorovici in \cite{MiltonNicorovici} for constant radial symmetric plasmonic structures in the two dimensional quasi-static regime. Their work has root from \cite{NicoroviciMcPhedranMilton94} (see also \cite{Nicorovici93, NicoroviciMcPhedranMiltonPodolskiy1}) where the localized resonance was observed and established for such a setting. 
Later, cloaking a source via ALR was studied by  Milton et al. in \cite{Milton-folded}, Bouchitte and Schweizer in \cite{BouchitteSchweizer10},  Ammari et al. in \cite{AmmariCiraoloKangLeeMilton, AmmariCiraoloKangLeeMilton2}, Kohn et al in \cite{KohnLu},, Nguyen and (H. L.) Nguyen  in \cite{MinhLoc1} in which special structures were considered due to the use of the separation of variables or the blow up of the power was investigated via spectral theory or variational method.  In \cite{Ng-CALR-CRAS, Ng-CALR, Ng-CALR-frequency}, Nguyen investigated cloaking a source via ALR  for a class of complementary media called  the class of  doubly complementary media for a general core-shell structure.  In these works,  the blow up of the power, the localized resonance,  and the cloaking effect are studied.  The reader can find a recent survey on the mathematical aspects for NIMs in \cite{Ng-Survey}.

\medskip 
In this paper, we add to the list of applications of NIMs  a new one namely cloaking {\bf an arbitrary object via ALR}. More precisely, we propose schemes for this type of cloaking and  provide the analysis for them.  Before stating the result, we introduce notations: 

\begin{notation} Given $R>0$ and $x \in \mR^d$, we denote $B(x, R)$  the open ball in $\mR^d$ centered at $x$ and  of radius $R$; when $x = 0$, we simply denote $B(x, R) $ by $B_R$. 
\end{notation}

The first result in this paper, whose proof is given in Section~\ref{sect-proof},  confirms that an annular plasmonic structure $-I$ cloaks small but finite size objects close to it  in  the quasistatic regime.  More precisely, we have

\begin{theorem}\label{thm-CS-1} Let $d=2$,  $0 <  r_0 <  r_1 < r_2 < R_0$, $x_1 \in \partial B_{r_1}$, and $x_2  \in \partial B_{r_2}$. Set  $r_3 = r_2^2/ r_1$ and  ${\cal C} := \big(B(x_1, r_0) \cap B_{r_1} \big) \cup  \big(B(x_2, r_0) \cap (B_{r_3}  \setminus B_{r_2}) \big)$ and let $a_c$ be a symmetric uniformly elliptic matrix-valued function defined in ${\cal C}$. 
Define 
\begin{equation}\label{def-AS-CS}
A_c = \left\{\begin{array}{cl} 
a_c &  \mbox{ in } {\cal C}, \\[6pt]
I & \mbox{ otherwise},
\end{array}\right. \quad 
\mbox{ and } \quad 
s_\delta = \left\{\begin{array}{cl}  -1 -  i \delta &  \mbox{ in } B_{r_2} \setminus B_{r_1}, \\[6pt]
1 & \mbox{ otherwise}.  
\end{array}\right. 
\end{equation} 
Given $f \in L^2(\mR^2)$ with $\supp f \subset B_{R_0} \setminus B_{r_3}$ and $\dsp \int_{\mR^2} f  = 0$, let  $u_\delta, \cU \in W^1(\mR^2)$ be respectively the unique solution to the equations
\begin{equation}\label{def-udelta-CS}
\dive(s_\delta A_c \nabla u_\delta )    = f \mbox{ in } \mR^2 
\end{equation}
and 
\begin{equation}\label{def-U-CS}
\Delta \cU     = f \mbox{ in } \mR^2. 
\end{equation}
For any $0 < \gamma < 1/2$, there exists $r_0(\gamma) > 0$,  which depends only on $r_1$ and $r_2$,  such that  if $r_0 < r_0(\gamma)$ then, for $R>0$,
\begin{equation}\label{result-CS}
\| u_\delta  -  \cU \|_{H^1(B_R \setminus B_{r_3})} \le C_R \delta^{\gamma} \| f\|_{L^2(\mR^2)}, 
\end{equation}
where $C_R$ is a positive constant independent of $f$, $\delta$, $r_0$, $x_1$, and $x_2$. 
\end{theorem}

\begin{figure}[h!]
\begin{center}
\includegraphics[width=10cm]{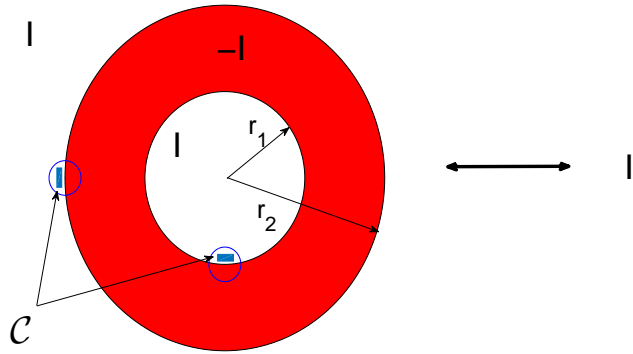} 
\caption{In the two dimensional quasistatic regime, the plamonic structure $-I$ in $B_{r_2} \setminus B_{r_1}$ (the red region) cloaks arbitrary objects located nearby in  ${\cal C}$. The medium on the left is equivalent to the homogeneous medium $I$ on the right for sources far away from the plasmonic structure.} \label{fig0}
\end{center}
\end{figure}

Here and in what follows, 
$$
W^1(\mR^2) : = \Big\{u \in H^1_{\loc}(\mR^2); \nabla u \in L^2(\mR^2) \mbox{ and } \frac{u(x)}{ \ln (2 + |x|) \sqrt{1 + |x|^2}}  \in L^2(\mR^2) \Big\}. 
$$

For an observer outside $B_{r_3}$, the medium $s_0A_c$ in $B_{r_3}$ looks like $I$ in $B_{r_3}$: the object $a_c$ in ${\cal C}$ is cloaked. No condition is imposed on $a_c$; any  symmetric uniformly elliptic matrix $a_c$ in ${\cal C}$ is allowed. The cloak $-I$ in $B_{r_2} \setminus B_{r_1}$ is {\bf independent} of the object.  
It is interesting to note that the plasmonic structure $-I$ in $B_{r_2} \setminus B_{r_1}$ can used in lens devices using NIMs, see \cite{NicoroviciMcPhedranMilton94, Ng-Superlensing}.

\medskip


In the two dimensional finite frequency regime, we obtain the following result whose proof is given in Section~\ref{sect-proof}.

\begin{theorem}\label{thm-CS-1-k}  Let $d=2$,  $0 < r_0 <  r_1 < r_2 < R_0$,  $x_1 \in \partial B_{r_1}$, and $x_2  \in \partial B_{r_2}$. Set   $r_3 = r_2^2/ r_1$  and  ${\cal C} := \big(B(x_1, r_0) \cap B_{r_1} \big) \cup  \big(B(x_2, r_0) \cap  (B_{r_3}  \setminus B_{r_2}) \big)$. Let $a_c$ be a symmetric uniformly elliptic matrix-valued function and let  
 $\sigma_c $ be a bounded real function both  defined in ${\cal C}$ such that  $a_c$ is piecewise $C^1$ and  $\sigma_c$ is bounded below by a positive constant. Define 
\begin{equation}\label{def-AS-CS-k}
\big( A_c, \Sigma_c \big) = \left\{\begin{array}{cl} 
\big(a_c, \sigma_c  \big) &  \mbox{ in } {\cal C}, \\[6pt]
\big( I, r_2^4/|x|^4 \big)& \mbox{ in } B_{r_2} \setminus B_{r_1}, \\[6pt]
\big( I, r_3^2/r_1^2 \big) & \mbox{ in } B_{r_1} \setminus {\cal C}, \\[6pt]
\big( I, 1\big) & \mbox{ otherwise}, 
\end{array}\right. 
\quad \mbox{ and } \quad  s_\delta = \left\{\begin{array}{cl}  -1 -  i \delta &  \mbox{ in } B_{r_2} \setminus B_{r_1}, \\[6pt]
1 & \mbox{ otherwise}.  
\end{array}\right. 
\end{equation} 
Given $f \in L^2(\mR^2)$ with $\supp f \subset B_{R_0} \setminus B_{r_3}$,  let $u_\delta, \cU \in H^1_{\loc}(\mR^2)$ be respectively the unique outgoing solution to the equations
\begin{equation}\label{def-udelta-CS-k}
\dive(s_\delta A_c \nabla u_\delta )  + k^2 s_0 \Sigma_c u_\delta   = f \mbox{ in } \mR^2
\end{equation}
and 
\begin{equation}\label{def-u-CS-k}
\Delta \cU + k^2 \cU    = f \mbox{ in } \mR^2. 
\end{equation}
For any $0 < \gamma < 1/2$, there exists $r_0 (\gamma)>0$,  which  depends only on $r_1$ and $r_2$ such that if $r_0  < r_0(\gamma)$ then  
\begin{equation}\label{result-CS-k}
\| u_\delta  - \cU\|_{H^1(B_{R} \setminus B_{r_3})}  \le C_R \delta^\gamma \| f\|_{L^2}, 
\end{equation}
for some positive constant $C_R$ independent of $f$, $\delta$, $r_0$, $x_1$, and $x_2$. 
\end{theorem}

Recall that a function $u \in H^1_{\loc}(\mR^d \setminus B_R)$ $(d \ge 2)$ for some $R>0$ which is a solution to the equation $\Delta u + k^2 u = 0 $ in $\mR^d \setminus B_R$ is said to satisfy the outgoing condition if 
\begin{equation*}
\partial_r u - i k u = o( r^{\frac{1-d}{2}}) \mbox{ as } r = |x| \to + \infty. 
\end{equation*}

For an observer outside $B_{r_3}$, the medium $(s_0A, s_0 \Sigma)$ in $B_{r_3}$ looks like $(I,1)$ in $B_{r_3}$: the object $(a_c, \sigma_c)$ in ${\cal C}$ is cloaked. No condition other than the standard ones is imposed on $a_c$ and $\sigma_c$. The cloak is {\bf independent} of the object.  It is interesting to note that the plasmonic structure $(-I, - r_2^4/|x|^4)$ in $B_{r_2} \setminus B_{r_1}$ can used in lens devices using NIMs, see \cite{PendryRamakrishna, Ng-Superlensing}.

\medskip 
In Section~\ref{sect-general},   we show  in two and three dimensions that for a class of doubly complementary media, introduced  in \cite{Ng-CALR, Ng-CALR-frequency} (see Definition~\ref{def-DCM} in Section~\ref{sect-general}), an arbitrary object can be disappeared if it is small and located close  to the plasmonic structure (see Theorem~\ref{thm-main-g} in Section~\ref{sect-general}). Therefore, a medium in  this class becomes a cloaking device which is {\bf independent} of the object.


\medskip 
Cloaking an object via ALR is related to but different from cloaking using complementary suggested in \cite{LaiChenZhangChanComplementary} and rigorously established in  \cite{Ng-Negative-Cloaking, MinhLoc2} for related schemes. Both types of cloaking use the concept of  complementary media to design cloaking devices. Nevertheless, in the cloaking using complementary media approach,  one cloaks an object by using its complementary to cancel the effect of light; hence  the cloaking device depends on the object.

\medskip 
Cloaking an object via ALR and cloaking a source via ALR share some similar figures but have some  different characters.  In the two dimensional quasistatic regime, one can use the plasmonic structure $-I$ in $B_{r_2} \setminus B_{r_1}$  as  a cloaking device in both settings. More general, doubly complementary media are used as a cloaking device in both types of cloaking (see \cite{Ng-CALR, Ng-CALR-frequency} and Section~\ref{sect-general}).  Nevertheless, concerning cloaking a source via ALR, the cloaking effect is {\bf relative} in the sense that the source is cloaked after being renormalized so that the power remains finite \footnote{The power is $\delta \| \nabla u_\delta\|_{L^2(B_{r_2} \setminus B_{r_1})^2}$.} (see e.g. \cite{Ng-CALR}). Concerning cloaking an object via ALR, the cloaking effect is {\bf not relative} in the sense that   one does not need to renormalize the power.  In fact, the power is finite in the setting of cloaking an object via ALR considered here (see Remarks~\ref{rem1-power} and \ref{rem2-power}). 

\medskip 
Milton and Nicorovici  in \cite{MiltonNicorovici} questioned that whether or not the structure $-I$ in $B_{r_2} \setminus B_{r_1}$ would cloak small objects nearby  in two dimensional quasistatic regime. Various numerical simulations on this effect were reported by Bruno and Lintner in  \cite{BrunoLintner07}; even though 
the cloaking effect is still mysterious and questionable. 
In this paper, the guest of Milton and Nicorovici was confirmed (Theorem~\ref{thm-CS-1}) and similar phenomena are observed and analyzed  in the finite frequency regime in both two and three dimensions (Theorems~\ref{thm-CS-1-k} and \ref{thm-main-g} ). As mentioned, the plasmonic structures used in Theorem~\ref{thm-CS-1} and  
\ref{thm-CS-1-k} can be used in lens devices using NIMs.   More generally,  it is established in \cite{Ng-Superlensing} that doubly complementary media can act like lenses. In this paper, we showed that they might become a cloak for  small but finite size objects near by. Thus the results presented here show that the modification given in \cite{Ng-Superlensing} from the suggestions in \cite{MiltonNicorovici, PendryRamakrishna} is necessary to ensure the superlensing property; otherwise, a lens can become a cloak (see Section~\ref{sect-CM-Superlensing}). In the same spirit, we as well show that it is necessary to modify the scheme 
of cloaking using complementary media suggested in \cite{LaiChenZhangChanComplementary} as done in \cite{Ng-Negative-Cloaking} (see Section~\ref{sect-CM-Cloaking}).

\medskip 
The analysis in this paper is on one hand based on the  use of reflecting and removing singularity techniques introduced in \cite{Ng-Superlensing, Ng-Negative-Cloaking} and on the other hand involves a new type of  three spheres inequality with ``partial data" and the use of conformal maps in two dimensions.

\medskip 

The rest of the paper is organized as follows. In Section~\ref{sect-general}, we present a scheme of cloaking an object via ALR for a class of doubly complementary media and provide their analysis. The main result of this section is Theorem~\ref{thm-main-g}.  In Section~\ref{sect-2d}, we first provide the proof of Theorems~\ref{thm-CS-1} and \ref{thm-CS-1-k} in Section~\ref{sect-proof}. We then show that the modified schemes considered in \cite{Ng-Superlensing} and \cite{Ng-Negative-Cloaking}  for superlensing and cloaking using complementary media are necessary in Sections \ref{sect-CM-Superlensing} and \ref{sect-CM-Cloaking}.

\section{A class of doubly complementary media acting as a cloaking device} \label{sect-general}

Let $k > 0$ and let $A$ be a real uniformly elliptic symmetric matrix-valued function   and $\Sigma$ be a  real  function bounded below and above by positive constants both defined in $\mR^d$ ($d = 2, 3$), i.e., $A$ is symmetric and for some $\Lambda \ge 1$ 
\begin{equation}\label{elliptic}
\Lambda^{-1} |\xi|^2 \le \langle A \xi, \xi \rangle \le \Lambda |\xi|^2 \quad \mbox{ and } \quad \Lambda^{-1} \le \Sigma \le \Lambda.
\end{equation}
Assume that  
\begin{equation}\label{cond-I}
A(x) = I,   \quad \Sigma(x) = 1  \mbox{ for large $|x|$},  
\end{equation}
and 
\begin{equation}\label{cond-C1}
A \mbox{ is piecewise } C^1.   
\end{equation}
The assumption \eqref{cond-C1} is used for the uniqueness of outgoing solutions. In what follows, in the case $k>0$, any matrix-valued function considered is assumed to satisfy \eqref{cond-C1}. 

\medskip 

Let $\Omega_1 \subset \subset \Omega_2 \subset \subset \mR^d$ be smooth  bounded simply connected open subsets of $\mR^d$,  and set, for $\delta \ge 0$,  
\begin{equation}\label{def-sd}
s_\delta (x) = \left\{\begin{array}{cl} - 1 - i \delta & \mbox{ in } \Omega_2 \setminus \Omega_1, \\[6pt]
1 & \mbox{ in } \mR^d \setminus (\Omega_2 \setminus \Omega_1). 
\end{array}\right.
\end{equation}
Given $f \in L^2(\mR^d)$ with compact support and   $\supp f \cap \Omega_2 = \O$,  and $\delta > 0$,  let  
$u_\delta \in H^1_{\loc}(\mR^d)$ be the unique outgoing  solution to 
\begin{equation}\label{def-ud}
\dive (s_\delta A \nabla u_\delta) + k^2 s_0 \Sigma u_\delta = f \mbox{ in } \mR^d. 
\end{equation}
In this section, we discuss the behavior of $u_\delta$ as $\delta \to 0$ when the medium inherits the doubly complementary property. To this end, 
we first recall the definition of reflecting complementary media and doubly complementary media introduced in \cite{Ng-Complementary} and \cite{Ng-CALR, Ng-CALR-frequency} respectively. 

\begin{definition}[Reflecting complementary media]   \fontfamily{m} \selectfont
 \label{def-Geo} Let $\Omega_1 \subset \subset \Omega_2 \subset \subset \Omega_3 \subset \subset \mR^d$ be smooth bounded  simply connected open subsets of $\mR^d$. The media $(A, \Sigma)$ in $\Omega_3 \setminus \Omega_2$ and $(-A, -\Sigma)$ in $\Omega_2 \setminus \Omega_1$  are said to be  {\it reflecting complementary} if 
there exists a diffeomorphism $F: \Omega_2 \setminus \bar \Omega_1 \to \Omega_3 \setminus \bar \Omega_2$ such that $F \in C^1(\bar \Omega_2 \setminus \Omega_1)$, 
\begin{equation}\label{cond-ASigma}
(F_*A, F_*\Sigma)   = (A, \Sigma )   \mbox{ for  } x \in  \Omega_3 \setminus \Omega_2, 
\end{equation}
\begin{equation}\label{cond-F-boundary}
F(x) = x \mbox{ on } \partial \Omega_2, 
\end{equation}
and the following two conditions hold: 1) There exists an diffeomorphism extension of $F$, which is still denoted by  $F$, from $\Omega_2 \setminus \{x_1\} \to \mR^d \setminus \bar \Omega_2$ for some $x_1 \in \Omega_1$; 2) There exists a diffeomorphism $G: \mR^d \setminus \bar \Omega_3 \to \Omega_3 \setminus \{x_1\}$ such that $G \in C^1(\mR^d \setminus \Omega_3)$, 
$G(x) = x \mbox{ on } \partial \Omega_3$, 
and $
G \circ F : \Omega_1  \to \Omega_3 \mbox{ is a diffeomorphism if one sets } G\circ F(x_1) = x_1.
$
\end{definition}

Here and in what follows, if ${\cal T}$ is a diffeomorphism and  $a$ and $\sigma$ are respectively a matrix-valued function and a complex function, we use the following standard notations  
\begin{equation}\label{def-F*}
{\cal T}_*a(y) = \frac{D{\cal T}(x) a(x) \nabla {\cal T}(x)^T}{|\det \nabla {\cal T}(x)|} \quad \mbox{ and } \quad {\cal T}_*\sigma(y) = \frac{ \sigma(x) }{|\det \nabla {\cal T}(x)|} \quad  \mbox{ where } x = {\cal T}^{-1}(y). 
\end{equation}

Conditions  \eqref{cond-ASigma} and \eqref{cond-F-boundary} are the main assumptions in Definition~\ref{def-Geo}. The key point behind this requirement is roughly speaking the following property:  if $u_0 \in H^1(\Omega_3 \setminus \Omega_1)$ is a solution of $\dive(s_0 A \nabla u_0) + k^2 s_0 \Sigma u_0= 0$ in $\Omega_3 \setminus \Omega_1$ and if $u_1$ is  defined in $\Omega_3 \setminus \Omega_2$ by  $u_1 = u_0 \circ F^{-1}$, then $\dive(A \nabla u_1) + k^2 \Sigma u_1 = 0$ in $\Omega_3 \setminus \Omega_2$, $u_1 - u_0 = A \nabla (u_1 - u_0) \cdot \nu  = 0 $ on $\partial \Omega_2$ (see e.g. Lemma~\ref{lem-TO}). Here and in what follows, $\nu$ denotes the outward unit vector on the boundary of a smooth bounded open subset of $\mR^d$. Hence $u_1 = u$ in $\Omega_3 \setminus \Omega_2$ by the unique continuation principle.   Conditions 1) and 2)  are  mild assumptions. Introducing $G$ makes the analysis more accessible, see \cite{Ng-Complementary, Ng-Superlensing, Ng-Negative-Cloaking, MinhLoc2} and the analysis presented in this paper.

%
%

\medskip 

We are ready to recall the definition of  doubly complementary media:  

\begin{definition} \label{def-DCM} \fontfamily{m} \selectfont The medium $(s_0 A, s_0\Sigma)$ is said to be {\it doubly complementary} if  for some $\Omega_2 \subset \subset \Omega_3$, $(A, \Sigma)$ in $\Omega_3 \setminus \Omega_2$ and $(-A, - \Sigma)$ in $\Omega_2 \setminus \Omega_1 $ are reflecting complementary, and 
\begin{equation}\label{def-DC}
F_*A = G_* F_*A = A  \quad \mbox{ and } \quad F_*\Sigma = G_* F_*\Sigma = \Sigma \mbox{ in } \Omega_3 \setminus \Omega_2, 
\end{equation}
for some $F$ and $G$ coming from Definition~\ref{def-Geo}. 
\end{definition}

\begin{figure}[h!]
\begin{center}
\includegraphics[width=6cm]{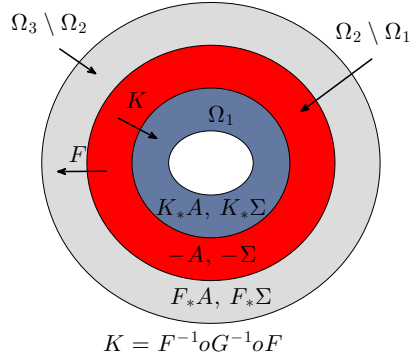} 
\caption{$(s_0A, s_0\Sigma)$ is doubly complementary: $(-A, -\Sigma)$ in $\Omega_2 \setminus \Omega_1$ (the red region) is complementary to $(F_*A, F_*\Sigma)$ in $\Omega_3 \setminus \Omega_2$ (the grey region) and $(K_*A, K_*\Sigma)$ with $K = F^{-1} \circ G^{-1} \circ F$ in $K(B_{r_2} \setminus B_{r_1})$ (the blue grey region).} \label{fig1}
\end{center}
\end{figure}

The reason for which media satisfying \eqref{def-DC} are called doubly complementary media is that $(-A, -\Sigma)$ in $B_{r_2} \setminus B_{r_1}$ is not only complementary to $(A, \Sigma)$ in $\Omega_3 \setminus \Omega_2$ but also to $(A, \Sigma)$ in $(G \circ F)^{-1}(\Omega_3 \setminus \overline{\Omega_2})$, a subset of $\Omega_1$ (see Figure \ref{fig1}). The key property behind Definition~\ref{def-DCM} is as follows. Assume that $u_0 \in H^1_{\loc}(\mR^d)$ is a solution of \eqref{def-ud} with $\delta = 0$ and $f = 0$ in $\Omega_3$. Set $u_1 = u_0 \circ F^{-1}$ and $u_{2} = u_1 \circ G^{-1}$. Then $u_0, \, u_1, \, u_2$ satisfy the  equation 
$\dive(A \nabla \cdot) + k^2 \Sigma \cdot = 0$  in $\Omega_3 \setminus \Omega_2$, $u_0 - u_1 = A \nabla u_0 \cdot \nu - A \nabla u_1 \cdot \nu = 0$ on $\partial \Omega_2$, and $u_1 - u_2 = A \nabla u_1 \cdot \nu - A \nabla u_2 \cdot \nu = 0$ on $\partial \Omega_3$ (see e.g. Lemma~\ref{lem-TO}). This implies $u_0 = u_1 = u_2$ in $\Omega_3 \setminus \Omega_2$.  

Taking $d=2$ and $r_3 = r_2^2/r_1$, and letting $F$ and $G$ be the Kelvin transform with respect to $\partial B_{r_2}$ and $\partial B_{r_3}$,  one can  verify that the media considered in Theorems~\ref{thm-CS-1} and \ref{thm-CS-1-k} with $r_0 = 0$ are of doubly complementary property.  
Theorems~\ref{thm-CS-1} and \ref{thm-CS-1-k} reveal that doubly complementary media might cloak small but finite size arbitrary objects nearby. 

\begin{remark} \label{rem-DCM} \fontfamily{m} \selectfont Given $(A, \Sigma)$ in $\mR^d$ and $\Omega_1 \subset \Omega_2 \subset \subset \mR^d$,  it is not easy in general to verify whether or not $(s_0A, s_0 \Sigma)$ is doubly complementary. Nevertheless, given $\Omega_1 \subset \Omega_2 \subset \subset \Omega_3 \subset \subset \mR^d$ and  $(A, \Sigma)$ in $\Omega_3 \setminus \Omega_2$, it is quite easy to choose $(A, \Sigma)$ in $\Omega_2$ such that $(s_0A, s_0\Sigma)$ is doubly complementary. One just needs to  choose diffeomorphisms $F$ and $G$ as in Definition~\ref{def-Geo} and define $(A, \Sigma) = (F^{-1}_*A, F^{-1}_*\Sigma)$ in $\Omega_2 \setminus \Omega_1$ and $(A, \Sigma) = (F^{-1}_*G^{-1}_*A, F^{-1}*G^{-1}_*\Sigma)$ in $F^{-1}\circ G^{-1} (\Omega_3 \setminus \Omega_2)$. 
\end{remark}

The following result established in  \cite[Theorem 1.1]{Ng-CALR-frequency} (see also \cite[Theorem 2 and Corollary 2]{Ng-Complementary}) provide an interesting property of doubly complementary media. 

\begin{proposition}\label{pro-recall}
Let  $d=2, 3$,  $k>0$, $0 < \delta < 1$, $f \in L^2(\mR^d)$  with compact support and  $\supp f \cap \Omega_3 = \O$,   and let $U_\delta \in H^1_{\loc}(\mR^d)$ be the unique outgoing solution of \eqref{def-ud}.  Assume that $(s_0A, s_0\Sigma)$ is doubly complementary. 
Then 
\begin{equation}\label{part2}
U_\delta \to \cU\mbox{ weakly in } H^1_{\loc}(\mR^d \setminus \Omega_3), 
\end{equation}
where $\cU \in H^1_{\loc}(\mR^d)$ is the unique outgoing solution of 
\begin{equation}\label{UU}
\dive (\cA \nabla {\cU}) + k^2 \cS \cU= f \mbox{ in } \mR^d. 
\end{equation}
Here 
\begin{equation}\label{def-cAS}
(\cA, \cS) := \left\{\begin{array}{cl} (A, \Sigma) & \mbox{ in } \mR^d \setminus \Omega_3, \\[6pt]
(G_*F_*A, G_*F_*\Sigma) & \mbox{ in } \Omega_3. 
\end{array}\right.
\end{equation}
\end{proposition}

The cloaking property for a class of doubly complementary media is given in the following 

\begin{theorem}\label{thm-main-g}
Let $d =2, \, 3$, $0< s < r_1 < 6 r_1 < r_2 < r_3 < R_0$. Set  
\begin{equation*}
H_{t} = \big\{x' \in \mR^{d-1}; \; |x'| < s \big\} \times [t , + \infty) \quad \mbox{ for } t\in \mR. 
\end{equation*} 
Assume that $(s_0A, s_0\Sigma)$ is doubly complementary 
with  $\Omega_1 = B_{r_1}$,  $\Omega_2 = B_{r_2} \setminus H_{2 r_1}$, and $\Omega_3 = B_{r_3}$. 
Let $a_c$ be a symmetric uniformly elliptic matrix-valued function  and $\sigma_c$ be a real  function bounded above and below by positive constants both defined in $H_{2r_1} \setminus H_{3 r_1}$. Assume that $A$ is $C^1$ in $B(z, r_1)$ with $z =(0, \cdots, 0, 4 r_1)$. Define 
\begin{equation*}
(A_{c}, \Sigma_c)  = \left\{ \begin{array}{cl}  (a_c, \sigma_c) & \mbox{ in } H_{2r_1} \setminus H_{3 r_1}, \\[6pt]
(A, \Sigma)  & \mbox{ otherwise}. 
\end{array}\right.
\end{equation*}
Let $k>0$, $0 < \delta < 1$, $f \in L^2(\mR^d)$ with $\supp f \subset B_{R_0 } \setminus B_{r_3}$   and let $u_{\delta} \in H^1_{\loc}(\mR^d)$ be the unique outgoing solution of 
\begin{equation}
\dive (s_\delta A_c \nabla u_\delta) + k^2 s_0 \Sigma_c u_\delta = f \mbox{ in } \mR^d. 
\end{equation}
For any $0 < \gamma < 1/2$, there exists a positive constant  $m $ depending only on $\gamma$, $r_1$, and the elliptic and the Lipschitz constant of $A$ in $B(z, r_1)$ such that if $r_1 > m s$ then 
\begin{equation}\label{part2-*}
\| u_{\delta} - \cU\|_{(B_R \setminus B_{r_3})} \le C_R \delta^\gamma \| f\|_{L^2}, 
\end{equation}
for some positive constant $C_R$ independent of $f$ and $\delta$  where $\cU \in H^1_{\loc}(\mR^d)$ is the unique outgoing solution of \eqref{UU}. 
\end{theorem}

\begin{figure}[h!]
\begin{center}
\includegraphics[width=6cm]{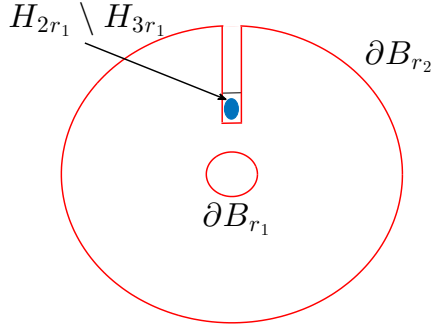} 
\caption{The cloaking device has the doubly complementary property and contains the plasmonic structure in the region $\Omega_2 \setminus B_{r_1}$ which has red boundary. The plamonic structure is  is charaterized by $(F^{-1}_*I, F^{-1}_*1)$ for some diffeomorphism $F:  \Omega_2 \setminus B_{r_1} \to B_{r_3} \setminus \bar \Omega_2$. The region $H_{2r_1} \setminus H_{3r_1}$ is the rectangle containing the blue object. Any object inside this region (for example the blue one) is cloaked.} \label{fig2}
\end{center}
\end{figure}

The geometry of the cloak is given in Figure~\ref{fig2}. 

\medskip

For an observer outside $B_{r_3}$, the medium $(s_0A, s_0 \Sigma)$ in $B_{r_3}$ looks like $(\hat A,\hat \Sigma)$ in $B_{r_3}$, which is independent of $(a_c, \sigma_c)$ in ${\cal C}$: the object $(a_c, \sigma_c)$ in ${\cal C}$ is cloaked. For example, by choosing the medium  $(A, \Sigma)$ in $\Omega_1$ in such a way that $G_*F_*A = I$ and $G_*F_*\Sigma = 1$ in $\Omega_3$, i.e., $A  = F^{-1}_*G^{-1}_*I$ and  $\Sigma  = F^{-1}_*G^{-1}_*1$ in $\Omega_1$  then $(\hat A, \hat \Sigma) = (I, 1)$: the object $(a_c, \sigma_c)$ in $H_{r_1} \setminus H_{2 r_1}$ disappeared in comparison with the homogeneous medium $(I, 1)$. 

\begin{remark} The assumption that $A$ is $C^1$ in $B(z, r_1)$ is required for the use of three spheres inequality. 
\end{remark}

\medskip 
The rest of this section containing two subsections is devoted to the proof of Theorem~\ref{thm-main-g}.  Some useful lemmas are presented in  the first subsection and 
the proof of Theorem~\ref{thm-main-g} is given in the second subsection. 

\subsection{Some useful lemmas}

We first recall a three spheres inequality which is an immediate consequence of  \cite[Theorem 2]{MinhLoc2}. 

\begin{lemma}\label{lem-MinhLoc}
	Let $d= 2, 3$, $c_1, c_2 >0$, $0 <   R_1  < R_2 <  R_3$, and let $a$ be  a Lipschitz  uniformly elliptic symmetric matrix-valued function defined in $B_{R_3}$, and $v \in H^1(B_{R_3} \setminus B_{R_1})$. Assume that 
	\begin{equation}\label{eq-Main}
|\dive (a \nabla v)| \le  c_1 |\nabla v|  + c_2|v|, \quad \mbox{ in } B_{R_3} \setminus B_{R_1}.
\end{equation}
There exists a constant $q \ge 1$, depending only on  the elliptic and the Lipschitz constants of $a$ such that
 \begin{equation}\label{def-alpha}
\| v\|_{\bH(\partial B_{R_2})} \le C\| v\|_{\bH(\partial B_{R_1})} ^{\alpha} \| v\|_{\bH(\partial B_{R_3})}^{1-\alpha}
\quad \mbox{ where } \quad 
	\alpha := \frac{R_2^{-q} - R_3^{-q}}{R_1^{-q} - R_{3}^{-q}}, 
\end{equation}
where,  for a smooth manifold $\Gamma$ of $\mR^d$ with or without boundary, 
\begin{equation}\label{def-norm-H}
\| v\|_{\bH(\Gamma)}: = \|v \|_{H^{1/2}(\Gamma)} + \|a \nabla v \cdot \nu  \|_{H^{-1/2}(\Gamma)} \; \;  \footnote{In this paper, $H^{-1/2}(\Gamma)$ denotes the duality of $H^{1/2}_0(\Gamma)$,  the completion of $C^1_{\mc}(\Gamma)$  in $H^{1/2}(\Gamma)$,  with the corresponding norm. Note that if $\Gamma$ has no boundary, then $H^{1/2}_0(\Gamma) = H^{1/2}(\Gamma)$.}. 
\end{equation}
Here $C$ is a positive constant  depending on the elliptic and the Lipschitz constants of $a$, and the constants $c_1,$ $c_2,$  $R_1, R_2, R_3$,  but independent of $v$. 
\end{lemma}

\begin{remark} In Lemma~\ref{lem-MinhLoc}, the constant $q$ is independent of $c_1$ and $c_2$ and there is no requirement on the smallness of $R_1, R_2$ and $R_3$. This is different from the standard three sphere inequalities obtained previously in the literature (see,  e.g.,  \cite{AlessandriniRondi}) and plays an important role in our analysis. 
\end{remark}

Applying Lemma~\ref{lem-MinhLoc}, we can establish 

\begin{lemma}\label{lem-3S-1} Let $d=2, 3$,  $0 < s < R$ and set $D = \big\{x' \in \mR^{d-1}; |x'| < s \big\} \times (-R, R)$. Let $a$ and $\sigma$ be respectively a Lipschitz  symmetric uniformly elliptic  matrix-valued function and a real bounded function both defined in $B_{R}$. Let  $f \in L^2(D)$ and $v \in H^1(D)$ be such that 
\begin{equation*}
\dive (a \nabla v) +  \sigma v = f \mbox{ in } D. 
\end{equation*}
Define $\Gamma = \big\{(x', x_d) \in \partial D \; ; \;  |x_d| \le  \frac{5}{6}R \big\}$.  For any $0 < \alpha < 1$, there exists a constant  $m >1$ depending only on the Lipschitz and the elliptic constants of $a$ such that if $R > m s$ then,  for some positive constant $C_{s, R}$ independent of $v$ and $f$, 
\begin{equation}\label{inequality-3S}
\| v\|_{H^1(B_{2 s} \cap D)} \le C_{s, R} \big(\| v \|_{\bH(\Gamma)} + \| f\|_{L^2} \big)^{\alpha} \big( \| v \|_{H^1(D)} + \| f\|_{L^2} \big)^{1 - \alpha}. 
\end{equation}
\end{lemma}

\begin{remark} Lemma~\ref{lem-3S-1} can be considered as a type of three sphrees inequality with ``partial data"
since no information of $v$ on $\partial D \setminus \Gamma$ is required. 
Lemma~\ref{lem-3S-1}  plays an important role in the proof of Theorem~\ref{thm-main-g}. 
\end{remark}

\noindent{\bf Proof.} We first consider the case $f = 0$. 
Fix $\varphi \in C^1_{\mc}(\mR^d)$ such that $\varphi = 1$ on $B_{2 R /3}$ and $\supp \varphi \subset B_{3R/4}$.  Let $w \in H^1(B_{4R/5} \setminus \Gamma)$ be such that 
\begin{equation*}
\dive (a \nabla w) +  \sigma w = 0 \mbox{ in } B_{4R/5} \setminus \Gamma, \quad a \nabla w \cdot \nu + i  w = 0 \mbox{ on } \partial B_{4 R/5} ,   
\end{equation*}
\begin{equation*}
[w] = \varphi v , \quad [a \nabla w \cdot \nu] = \varphi a \nabla v \cdot \nu \mbox{ on } \Gamma. 
\end{equation*}
Here and in what follows, on $\partial D$,  $[ u ] : = u |_{\mR^d \setminus D} - u |_{D}$ for an appropriate function $u$; similar notation is used for  $[a \nabla u \cdot \nu]$ on $\partial D$.   
Since  
$$
\| \varphi v \|_{H^{1/2}(\Gamma)} + \| \varphi a \nabla v \cdot \nu  \|_{H^{-1/2}(\Gamma)} \le C \big( \|v \|_{H^{1/2}(\Gamma)} + \| a \nabla v \cdot \nu  \|_{H^{-1/2}(\Gamma)}\big), 
$$
it follows that 
\begin{equation}\label{lem-3S-1-part1}
\| w\|_{H^1(B_{4 R /5} \setminus \Gamma)} \le C \| v \|_{\bH(\Gamma)}.  
\end{equation}
Here and in what follows $C$ denotes a positive constant independent of $v$. 
Define 
\begin{equation}\label{lem-3S-1-part2}
V = \left\{ \begin{array}{cl}  w + v & \mbox{ in } D, \\[6pt]
w & \mbox{ in } B_{4R/5} \setminus D. 
\end{array}\right.
\end{equation}
Note that 
\begin{equation}\label{rem-toto}
\| v \|_{\bH(\Gamma)} \le C_{s, R} \| v \|_{H^1(D)}. 
\end{equation}
Since $\varphi = 1$ in $B_{2R/3}$, it follows from  \eqref{lem-3S-1-part1}, \eqref{lem-3S-1-part2}, and \eqref{rem-toto} that $V \in H^1(B_{2 R/3 })$, 
\begin{equation}\label{lem-3S-1-part3}
\|V \|_{H^1(B_{2 R / 3})} \le C \| v\|_{H^1(D)}, 
\end{equation}
and 
\begin{equation}\label{eq-V}
\dive (a \nabla V) + \sigma V = 0 \mbox{ in } B_{2R/3}. 
\end{equation}
Set $z = (R/\sqrt{m}, 0, \cdots, 0)$, $R_1 = (R/ \sqrt{m}) - s$, $R_2 = (R/\sqrt{m}) + 4s$, and $R_3 = R/3$.  Applying Lemma~\ref{lem-MinhLoc}, we have 
\begin{equation}\label{lem-3S-1-part4}
\| V\|_{ \bH \big(\partial B(z, R_2) \big)} \le C \| V\|_{\bH \big(\partial B(z, R_1) \big)}^{\alpha} \|V \|_{\bH \big(\partial B(z, R_3) \big)}^{1 - \alpha}, 
\end{equation}
with 
\begin{equation}\label{def-alpha-1}
\alpha= \frac{R_2^{-q} - R_3^{-q}}{R_1^{-q} - R_3^{-q}}, 
\end{equation}
for some $q>1$ independent of $v$, $R$, $m$, and $s$. From \eqref{eq-V}, we derive from Lemma~\ref{lem-compact} below that, for some positive constant $C>1$,  
\begin{equation}\label{lem-3S-1-part4-1}
C^{-1} \| V\|_{ H^1 \big(B(z, r) \big)} \le  \| V\|_{ \bH \big(\partial B(z, r) \big)} \le C \| V\|_{ H^1 \big(B(z, r) \big)} 
\end{equation}
with $r = R_1, \; R_2,$ or $R_3$. It follows  from \eqref{lem-3S-1-part4} that 
\begin{equation}\label{lem-3S-1-part5}
\| V\|_{ H^1 \big(B(z, R_2) \big)} \le C \| V\|_{H^1 \big(B(z, R_1) \big)}^{\alpha} \|V \|_{H^1 \big(B(z, R_3) \big)}^{1 - \alpha}. 
\end{equation}
Note that   $\alpha$ can be chosen arbitraly close to $1$ by taking $R> ms$ and large $m$ in \eqref{def-alpha-1}. 
The conclusion in the case $f = 0$ follows from \eqref{lem-3S-1-part1}, \eqref{lem-3S-1-part3}, and \eqref{lem-3S-1-part5} by the definition of $V$ in \eqref{lem-3S-1-part2}; in particular, $V = w$ in $B(z, R_1)$.

\medskip 
We next consider the case $f \in L^2(D)$. Let $v_1 \in H^1(D)$ be such that 
$$
\dive(a \nabla v_1) + \sigma v_1 = f \mbox{ in } D \quad \mbox{ and } \quad  a \nabla  v_1  \cdot \nu - i v_1 = 0 \mbox{ on } \partial D. 
$$
Then 
$$
\| v_1\|_{H^1(D)} \le C \|f \|_{L^2} \quad \mbox{ and } \quad \| v_1\|_{\bH(\partial D)} \le C \| f\|_{L^2(D)}. 
$$
The conclusion in the general case now follows from the case $f = 0$ by applying the previous case for $v - v_1$. The proof is complete. \proofend

\medskip 
In the proof of Lemma~\ref{lem-3S-1}, we use the following result 
\begin{lemma}\label{lem-compact}
Let $d=2, 3$, $R>0$, $a$ and $\sigma$ be respectively a Lipschitz  symmetric uniformly elliptic  matrix-valued function and a real bounded function both defined in $B_R$. Let  $v \in H^1(B_R)$ be such that 
\begin{equation*}
\dive (a \nabla v) +  \sigma v = 0 \mbox{ in } B_R. 
\end{equation*}
Then, for some positive constant $C > 1$, independent of $v$, 
\begin{equation}\label{lem-compact-state1}
C^{-1} \| v\|_{ H^1 (B_R )} \le  \| v\|_{ \bH (\partial B_R)} \le C \| v\|_{ H^1 (B_R)} 
\end{equation}
\end{lemma}

\noindent{\bf Proof.}  The second inequality of \eqref{lem-compact-state1} is a consequence of the trace theory. It remains to establish the first inequality of \eqref{lem-compact-state1}. We first prove that 
\begin{equation}\label{lem-compact-contradiction}
\| v\|_{L^2(B_R)} \le C \| v\|_{ \bH (\partial B_R)}
\end{equation}
by contradiction.  In this proof, $C$ denotes a positive constant independent of $v$ and $n$. 
Suppose that it is not true. There exists some sequence $(v_n) \subset H^1(B_R)$ such that 
\begin{equation}\label{lem-compact-state2}
\dive (a \nabla v_n) +  \sigma v_n = 0 \mbox{ in } B_R 
\end{equation}
and 
\begin{equation}\label{lem-compact-state3}
n \| v_n\|_{ \bH (\partial B_R)} \le \| v_n\|_{ L^2 (B_R )} = 1. 
\end{equation}
Multiplying the equation of $v_n$ \eqref{lem-compact-state2} by $\bar v_n$ (the conjugate of $v_n$), integrating in $B_R$, and using \eqref{lem-compact-state3}, we obtain 
\begin{equation*}
\int_{B_R} |\nabla v_n|^2 \le C \int_{B_R} |v_n|^2 + C. 
\end{equation*}
Thus $(v_n)$ is bounded in $H^1(B_R)$. Without loss of generality, one may assume that $(v_n)$ converges to $v$ weakly in $H^1(B_R)$ and strongly in $L^2(B_R)$ to $v$ for some $v \in H^1(B_R)$. It is clear from \eqref{lem-compact-state2} and \eqref{lem-compact-state3} that 
\begin{equation*}
\dive (a \nabla v) +  \sigma v = 0 \mbox{ in } B_R \quad \mbox{ and } \quad \|v \|_{\bH(\partial B_R)} = 0. 
\end{equation*}
By the unique continuation principle, we have $v = 0 \mbox{ in } B_R.
$
This contradicts the fact $\| v\|_{L^2(B_R)}  = \lim_{n \to + \infty} \| v_n\|_{L^2(B_R)} = 1$. Hence \eqref{lem-compact-contradiction} is established. 

\medskip 
The first inequality of \eqref{lem-compact-state1} is now a consequence of \eqref{lem-compact-contradiction} after multiplying the equation of $v$ by $\bar v$ and integrating in $B_R$. \proofend

 \medskip 
We next recall a stability result of \eqref{def-ud} 
 established in \cite[Lemma 2.1]{Ng-CALR-frequency}.  

\begin{lemma}\label{lem1} Let $d =2, 3$, $k>0$, $\delta_0 > 0$,  $R_0> 0$, $g \in L^2(\mR^d)$ with support in $B_{R_0}$. For $0 < \delta < \delta_0$, there exists a unique outgoing  solution $v_\delta \in H^1_{\loc}(\mR^d)$ of \eqref{def-ud}. Moreover,
\begin{equation}\label{stability}
\| v_\delta \|_{H^1(B_R)}^2 \le  \frac{C_R}{\delta} \Big| \int g \bar v_\delta \Big| + C_R  \| g\|_{L^2}^2, 
\end{equation}
for some positive constant $C_R$ independent of $g$ and $\delta$.
\end{lemma}

We end this section by recalling the following change of variables formula \cite[Lemma 2]{Ng-Complementary} which is used several times  in this paper. 

\begin{lemma}\label{lem-TO} Let $D_1 \subset \subset D_2 \subset \subset D_3$  be three smooth bounded open subsets of $\mR^d$. Let $a \in [L^\infty(D_2 \setminus D_1)]^{d \times d}$,  $\sigma \in L^\infty(D_2 \setminus D_1)$ and let  ${\cal T}$ be a diffeomorphism from $D_2 \setminus \bar D_1$ onto $D_3 \setminus \bar D_2$. Assume that   $u \in H^1(D_2 \setminus D_1)$ and set $v = u \circ {\cal T}^{-1}$. Then
\begin{equation*}
\dive (a \nabla u) +  \sigma u = f \mbox{ in } D_2 \setminus D_1, 
\end{equation*}
for some $f \in L^2(D_2 \setminus D_1)$,  if and only if
\begin{equation}\label{TO-eq}
\dive ({\cal T}_*a \nabla v) + {\cal T}_*\sigma v = {\cal T}_* f \mbox{ in } D_3 \setminus D_2. 
\end{equation}
Assume in addition that ${\cal T}(x) = x$ on $\partial D_2$. Then  
\begin{equation}\label{TO-bdry}
v = u \quad \mbox{ and } \quad {\cal T}_*a \nabla v \cdot \nu = - a \nabla u \cdot \nu  \mbox{ on } \partial D_2.
\end{equation}
\end{lemma}


\subsection{Proof of Theorem~\ref{thm-main-g}}

The proof uses the reflecting and removing localized singularity techniques introduced in \cite{Ng-Superlensing, Ng-Negative-Cloaking}. 
Applying Lemma~\ref{lem1}, we have 
\begin{equation}\label{part0-thm-main-g}
\| u_{\delta} \|_{H^1(B_R)}\le C \Big(\delta^{-1/2} \|f \|_{L^2}^{1/2} \| u_\delta \|_{L^2(\supp f)}^{1/2} + \| f\|_{L^2}\Big). 
\end{equation} 
Here and in what follows in this proof, $C$ denotes a positive constant independent of $\delta$ and $f$. 
Set 
$$
S  = H_{2 r_1} \setminus H_{3 r_1}. 
$$
Define 
\begin{equation*}
u_{1, \delta} = u_\delta \circ F^{-1} \mbox{ in } \mR^d \setminus \Omega_2 \quad \mbox{ and } \quad 
u_{2, \delta} = u_{1, \delta} \circ G^{-1} \mbox{ in } B_{r_3}, 
\end{equation*}
where $F$ and $G$ come from the definition of doubly complementary media. 
Applying Lemma~\ref{lem-TO}, we obtain,  since $G_*F_*A = F_*A = A $ in $B_{r_3} \setminus \Omega_2$ (the medium is of doubly complementary property)
\begin{equation}\label{transmission1-thm-main-g}
u_{1, \delta} = u_\delta, \quad (1 + i \delta) A  \nabla  u_{1, \delta} \cdot \nu = A \nabla  u_{\delta}  \big|_{B_{r_3} \setminus \Omega_2}\cdot \nu \mbox{ on } \partial \Omega_2 \setminus \partial S,  
\end{equation}
\begin{equation}\label{transmission2-thm-main-g}
u_{2, \delta} = u_{1, \delta}, \quad  \mbox{ and }  \quad A \nabla  u_{2, \delta} = (1 + i \delta) A \nabla  u_{1, \delta}  \big|_{B_{r_3} \setminus \Omega_2} \cdot \nu \mbox{ on } \partial  B_{r_3}. 
\end{equation}
This implies 
\begin{equation}\label{transmission1-thm-main-g-1}
\|u_{1, \delta} - u_{\delta} \|_{\bH(\partial \Omega_2 \setminus \partial S)} \le C \delta \| u_\delta\|_{H^1(B_{r_3})} 
\end{equation}
and
\begin{equation}\label{transmission2-thm-main-g-2}
\|u_{2, \delta} - u_{1, \delta} \|_{\bH(\partial B_{r_3})} \le C \delta \| u_\delta\|_{H^1(B_{r_3})}.
\end{equation}
Applying Lemma~\ref{lem-TO}, we have, since $F_*A = A$ and $F_*\Sigma = \Sigma$ in $B_{r_3} \setminus \Omega_2$ and $f = 0$ in $B_{r_3}$,  
$$
\dive\big((1 + i \delta) A \nabla u_{1, \delta} \big) + k^2 \Sigma u_{1, \delta}= 0 \mbox{ in } B_{r_3} \setminus \Omega_2; 
$$
which yields 
\begin{equation}\label{haha11}
\dive(A \nabla u_{1, \delta}) + k^2 \Sigma u_{1, \delta}= \Big(1 - \frac{1}{1 + i \delta} \Big) k^2 \Sigma u_{1, \delta} = - \frac{i \delta}{1 + i \delta } k^2 \Sigma u_{1, \delta} \mbox{ in } B_{r_3} \setminus \Omega_2. 
\end{equation}
Recall that 
\begin{equation}\label{haha12}
\dive(A \nabla u_{1, \delta}) + k^2 \Sigma u_{1, \delta}= 0 \mbox{ in } (B_{r_3} \setminus \Omega_2) \setminus S. 
\end{equation}
Set $\alpha = \gamma + 1/ 2 < 1$. Applying Lemma~\ref{lem-3S-1} with $D = \big\{x' \in \mR^{d-1}; |x'| < s \big\} \times (-r_1, r_1)$ and $v(x) = u_{1, \delta}(x + z) - u_\delta (x + z)$ for $x \in D$, we derive from \eqref{transmission1-thm-main-g-1}, \eqref{haha11}, and \eqref{haha12} that if $r_1 > m s$ and $m$ is sufficiently large then
\begin{equation}\label{claim1-thm-main-g}
\|u_{1, \delta} - u_{\delta}\|_{H^1\big((H_{3r_1}  \setminus H_{5r_1} )\cap B(z, 2s) \big)} \le C \delta^{\alpha} \| u_\delta \|_{H^1(B_{r_3})}. 
\end{equation} 
Define  $O_2 = B_{r_2} \setminus H_{4r_1}$ and $O = B_{r_3 } \setminus O_2$,  and set 
\begin{equation*}
{\cal U}_\delta  = \left\{\begin{array}{cl} u_\delta & \mbox{ in } \mR^d \setminus B_{r_3}, \\[6pt]
u_{2, \delta} + u_\delta -   u_{1, \delta}  & \mbox{ in } O, \\[6pt]
u_{2, \delta} &  \mbox{ in } O_2. 
\end{array}\right. 
\end{equation*}
Then ${\cal U}_\delta \in H^1 \big(B_R\setminus \partial O \big)$ for all $R> 0$ and $U_\delta$ is an outgoing solution of the equation
\begin{equation*}
\dive (\hat A \nabla {\cal U}_\delta) + k^2 \hat \Sigma  {\cal U}_{\delta} = f \mbox{ in } \mR^d \setminus  \partial O.  
\end{equation*}
Note that $\hat A$ is uniformly elliptic and  $\hat \Sigma$  is bounded above and below by positive constants. It follows that 
\begin{equation}\label{thm3-part1-1}
\|{\cal U}_\delta \|_{H^1(B_R \setminus \partial O)} \\[6pt] 
\le C_R \Big( \|[{\cal U}_\delta] \|_{H^{1/2}(\partial O)} + \|[\hat A \nabla  {\cal U}_\delta  \cdot \nu] \|_{H^{-1/2}(\partial O)} \Big). 
\end{equation}
By the definition of ${\cal U}_\delta$, we have
\begin{equation*}
\|[{\cal U}_\delta] \|_{H^{1/2}(\partial O_2)} + \|[\hat A \nabla  {\cal U}_\delta  \cdot \nu] \|_{H^{-1/2}(\partial O_2)} =  \|u_\delta - u_{1, \delta} \|_{H^{1/2}(\partial O_2)} +  \|A \nabla  (u_\delta - u_{1, \delta})  \cdot \nu\|_{H^{-1/2}(\partial O_2)}
\end{equation*}
and
\begin{multline*}
\|[{\cal U}_\delta] \|_{H^{1/2}(\partial B_{r_3})} + \|[\hat A \nabla  {\cal U}_\delta  \cdot \nu] \|_{H^{-1/2}(\partial B_{r_3})} \\[6pt]
=  \|u_{2, \delta} - u_{1, \delta} \|_{H^{1/2}(\partial B_{r_3})} +  \|A \nabla  (u_{2, \delta} - u_{1, \delta})  \cdot \nu\|_{H^{-1/2}(\partial B_{r_3})}. 
\end{multline*}
Since $\partial O = \partial B_{r_3} \cup \partial O_2$,  we derive from \eqref{transmission1-thm-main-g-1}, \eqref{transmission2-thm-main-g-2}, and  \eqref{claim1-thm-main-g} that 
\begin{equation}\label{thm3-par1-1-1}
\|[{\cal U}_\delta] \|_{H^{1/2}(\partial O)} + \|[\hat A \nabla {\cal U}_\delta  \cdot \nu] \|_{H^{-1/2}(\partial O)} \le  C \delta^{\alpha} \| u_\delta \|_{H^1(B_{r_3})}. 
\end{equation}
It follows from  \eqref{part0-thm-main-g} and \eqref{thm3-par1-1-1} that, for $R> R_0$, 
\begin{equation*}
\| {\cal U}_\delta\|_{H^1(B_{R} \setminus \partial O)} \le  C_R \delta^{\alpha} \Big( \delta^{-1/2}\| {\cal U}_\delta \|_{L^2(B_R \setminus B_{r_3})}^{1/2} \|f \|_{L^2}^{1/2} + \| f\|_{L^2}\Big) + C_R \|f \|_{L^2}.
\end{equation*}
since $\supp f \subset B_R \setminus B_{r_3}$. 
Since $\alpha = \gamma + 1/2  > 1/2$,  it follows that 
\begin{equation}\label{thm3-part1-1-2}
\| {\cal U}_\delta\|_{H^1(B_{R} \setminus \partial O)} \le  C_R  \|f \|_{L^2}.
\end{equation}
Since $\supp f \cap B_{r_3}  = \O$, we derive from \eqref{part0-thm-main-g} that, for $R>0$,  
\begin{equation}\label{ud-improve}
\| u_\delta\|_{H^1(B_R)} \le C_R \delta^{-1/2} \|f \|_{L^2}; 
\end{equation}
which yields, by \eqref{thm3-par1-1-1},  
\begin{equation}\label{haha-thm3}
 \|[{\cal U}_\delta] \|_{H^{1/2}(\partial O)} + \|[\hat A \nabla  {\cal U}_\delta  \cdot \nu] \|_{H^{-1/2}(\partial O)} \le C \delta^{\alpha - 1/ 2} \| f\|_{L^2}. 
\end{equation}
Hence ${\cal U}_\delta$ is bounded in $H^1 (B_R \setminus  \partial  O )$. 
Without loss of generality,  one may assume that ${\cal U}_{\delta} \to {\cal U}$ weakly in $H^1 (B_R \setminus  \partial  O) $  as $\delta \to 0$ for any $R>0$; moreover, ${\cal U} \in H^1_{\loc}(\mR^d)$  is the unique outgoing solution to the equation  
\begin{equation*}
\dive (\hat A \nabla  {\cal U}) + k^2 \hat \Sigma  {\cal U} = f \mbox{ in } \mR^d. 
\end{equation*}
Hence ${\cal U} = \cU$ in $\mR^d$. Since the limit is unique,  the convergence holds for the whole family $({\cal U}_{\delta})$ as $\delta \to 0$. In other words,  ${\cal U}_\delta \to \cU$ in $H^1_{\loc}(\mR^d)$. 

\medskip 
To obtain the rate of convergence, let us consider the equation of $U_\delta - \hat u$. We have 
\begin{equation*}
\dive \big(\hat A \nabla ( {\cal U}_\delta - \hat u) \big) + k^2 \hat \Sigma  ({\cal U}_{\delta}- \hat u) = 0\mbox{ in } \mR^d \setminus \partial O.  
\end{equation*}
As in \eqref{thm3-part1-1}, we have
\begin{equation}\label{thm3-part2-1}
\|{\cal U}_\delta -\hat u \|_{H^1 (B_R \setminus \partial O )} \\[6pt] 
\le C_R \Big( \|[{\cal U}_\delta ] \|_{H^{1/2}(\partial O)} + \|[\hat A \nabla  {\cal U}_\delta  \cdot \nu] \|_{H^{-1/2}(\partial O)} \Big). 
\end{equation}
The conclusion now follows from \eqref{haha-thm3} and \eqref{thm3-part2-1}. \proofend

\begin{remark}\label{rem1-power} The power of $u_\delta$ defined by $\dsp \int_{\Omega_2 \setminus \Omega_1} \delta |\nabla u_\delta|^2$
is finite in the setting considered in Theorem~\ref{thm-main-g} by \eqref{ud-improve}. 
\end{remark}

\begin{remark} It is showed in \cite{Ng-WP} that resonance takes place for reflecting complementary media. Various conditions on the stability of the Helmholtz equations with sign changing coefficients were given there and the necessity of the reflecting complementary property of media for the occurrence of the resonance was discussed. 
\end{remark}

\section{Cloaking an object via ALR in two dimensions and related problems} \label{sect-2d}

This section containing three subsections is  devoted to cloaking an object via ALR in two dimensions and related problems. 
In the first section,  we present the proofs of  Theorems~\ref{thm-CS-1} and \ref{thm-CS-1-k}.
In the second section, we make some comments on superlensing using complementary media. More precisely, we show that a modifications as in \cite{Ng-Superlensing} from the suggestion  in \cite{NicoroviciMcPhedranMilton94, PendryNegative, PendryCylindricalLenses} is necessary: without the modification, a lens can become a cloaking device (see Proposition~\ref{pro-lensing}).  Finally, we make some comments on cloaking using complementary media. We  prove that without the modification suggested in \cite{Ng-Negative-Cloaking}, cloaking effect might not be achieved for the schemes proposed  in    \cite{LaiChenZhangChanComplementary} (see Proposition~\ref{pro-cloaking}). 

\medskip 

In this section, we  use the complex notations for several places, for example, the polar coordinate $z = r e^{i \theta}$ is used and  for $x \in \mR$,  $B(x, R)$ means $B(z, R)$ with $z = x$.

\subsection{Proofs of Theorems~\ref{thm-CS-1} and \ref{thm-CS-1-k}} \label{sect-proof}

We first state a  variant of Lemma~\ref{lem-3S-1}.

\begin{lemma}\label{lem-3S-2} Let $0< R \le R_1$, $m \in \mN$,  and define $D = T \big(B_{R} \setminus B(-R_1, R_1) \big)$  where $T(z) = z^{1/m}$  \;  \footnote{The following definition of $T(z)$ is used in this paper: for $z = r e^{i \theta}$ with $r \ge 0$, $\theta \in (-\pi, \pi)$, we define $T(z) = r^{1/m} e^{i \theta/m}$.}  . 
Set $\hat D =  \big\{ z \in  D; \; \frac{1}{20}R^{1/m} < |z| < R^{1/m} \big\}$ and $\Gamma = \big\{ z \in \partial D; \; \frac{1}{10}R^{1/m} < |z| < \frac{9}{10}R^{1/m} \big\}$. Let  $\sigma$ be a real bounded function defined in $\hat D$, $f \in L^2(\hat D)$, and  $v \in H^1(\hat D)$ be such that 
\begin{equation*}
\Delta v +  \sigma v = f \mbox{ in } \hat D. 
\end{equation*}
For any $0 < \alpha < 1$, there exists $m_0 \in \mN$ depending only on $R$ such that if $m > m_0$,
then,  for some neighborhood $D_1$ of $\big\{z \in D; \; |z| = \frac{1}{2} R^{1/m} \big\}$ and  for some positive constant $C$ both independent of $v$ and $f$, 
\begin{equation*}
\| v\|_{H^1(D_1)} \le C  \big(\| v \|_{\bH(\Gamma)} + \| f\|_{L^2(\hat D)} \big)^{\alpha} \big( \| v \|_{H^1(\hat D)} + \| f\|_{L^2(\hat D)} \big)^{1 - \alpha} .  
\end{equation*}
\end{lemma}

\noindent{\bf Proof.} The proof of Lemma~\ref{lem-3S-2} is similar to the one of Lemma \ref{lem-3S-1}. For the convenience of the reader, we present the proof. We first consider the case $f = 0$ in $\hat D$. 
For the simplicity of notations, set $\hR = R^{1/m}$. 
We first assume that $m$ is sufficiently large such that $1/2 < \hR< 2$; this is possible since $\lim_{m \to + \infty} R^{1/m} = 1$.   Fix $\varphi \in C^1(\mR^2)$ such that $\varphi = 1$ in $B_{\hR /4}$  and $\supp \varphi \subset B_{7\hR/24}$,  and define $\varphi_m(z) = \varphi(z - \hR/2)$.  Set $O = B(\hR/2, \hR/3)$, the disk centered at $\hR/2$ and of radius $\hR/3$,  and extend $\sigma$ by $1$ in $O \setminus \hat D$. We  still denote the extension of $\sigma$ by $\sigma$ for simplicity of notations. Let  $w \in H^1(O\setminus \Gamma)$ be the unique solution of the system 
\begin{equation*}
\Delta  w +  \sigma w = 0 \mbox{ in } O \setminus \Gamma, \quad \partial_\nu w -  i  w = 0 \mbox{ on } \partial O,   
\end{equation*}
\begin{equation*}
[w] = \varphi_m v , \quad \mbox{ and } \quad  [\partial_\nu w ] = \varphi_m \partial_\nu v  \mbox{ on } \Gamma. 
\end{equation*}
Since
$$
\|\varphi_m v\|_{H^{1/2}(\Gamma)} + \|\varphi_m  \partial_\nu v  \|_{H^{-1/2}(\Gamma)} \le C \big(\| v\|_{H^{1/2}(\Gamma)} + \|\partial_\nu v \|_{H^{-1/2}(\Gamma)}\big),
$$
it follows that 
\begin{equation}\label{lem-3S-2-part1}
\| w\|_{H^1(O \setminus \Gamma)} \le C \| v \|_{\bH(\Gamma)}.  
\end{equation}
Here and in what follows $C$ denotes a positive constant independent of $v$. 
Define 
\begin{equation}\label{lem-3S-2-part2}
V = \left\{ \begin{array}{cl}  w + v & \mbox{ in } \hat D, \\[6pt]
w & \mbox{ in } O \setminus \hat D. 
\end{array}\right.
\end{equation}
Then $V \in H^1(O)$, 
$$
\Delta V+ \sigma V = 0 \mbox{ in } O, 
$$
 and 
\begin{equation}\label{lem-3S-2-part3}
\|V \|_{H^1(O)} \le C \| v\|_{H^1(\hat D)}. 
\end{equation}
Set $z =\hR/2 + i \hR / \sqrt{m}$, $R_1 = \hR/ \sqrt{m} - 2 \pi \hR/m$, $R_2 = \hR/ \sqrt{m} + 2 \pi \hR/ m$, and $R_3 = \hR/4$.  Applying Lemma~\ref{lem-MinhLoc}, we have 
\begin{equation}\label{lem-3S-2-part4}
\| V\|_{H^1\big(B(z, R_2) \big)} \le C \| V\|_{H^1\big(B(z, R_1) \big)}^{\alpha} \|V \|_{H^1\big(B(z, R_3) \big)}^{1 - \alpha}, 
\end{equation}
where 
\begin{equation*}
\alpha= \frac{R_2^{-q} - R_3^{-q}}{R_1^{-q} - R_3^{-q}},
\end{equation*}
for some $q>1$ independent of $v$, $\hR$, and $m$. Here we use the fact that, by Lemma~\ref{lem-compact}, 
\begin{equation*}
C^{-1} \| V\|_{ H^1 \big(B(z, r) \big)} \le  \| V\|_{ \bH \big(\partial B(z, r) \big)} \le C \| V\|_{ H^1 \big(B(z, r) \big)} 
\end{equation*}
with $r = R_1, R_2$,  or $R_3$.  The conclusion in the case $f = 0$ follows from \eqref{lem-3S-2-part1}, \eqref{lem-3S-2-part3}, \eqref{lem-3S-2-part4},  and the definition of $V$ in \eqref{lem-3S-2-part2} by noting that $B(z, R_1) \subset O \setminus \hat D$ (hence $V = w$ in $B(z, R_1)$),  $B(z, R_2)$ contains a neighborhood of $\{ z \in D; \; |z| = \hR/2\}$, and $\alpha$ can be taken aribitrary close to 1 if $m$ is large enough.  

\medskip 
The conclusion in the general case follows from the case $f = 0$ by applying the result in the case $f = 0$ for $v - v_1$ where $v_1 \in H^1(\hat D)$ is the unique solution of the system 
$$
\Delta v_1 + \sigma v_1 = f \mbox{ in } \hat D \quad \mbox{ and } \quad \partial_{\nu} v_1 - i v_1 = 0 \mbox{ on } \partial \hat D. 
$$
The proof is complete.  \proofend

\medskip 
As a consequence of Lemma~\ref{lem-3S-1}, we have 
\begin{corollary}\label{cor1}
Let  $0 < r <  R \le R_1$, and define $D = B_R \setminus B(-R_1, R_1)$. Set $\widetilde D =  \big\{ z \in D; r < |z| <  R \big\}$ and $\Gamma = \big\{ z \in \partial D; \; \frac{1}{10}R < |z| < \frac{9}{10}R\big\}$. Let    $\sigma$ be a real bounded function defined in $\widetilde D$, $f \in L^2(\widetilde D)$, and   $v \in H^1(\widetilde D)$ be such that 
\begin{equation*}
\Delta v +  \sigma v = f \mbox{ in } \widetilde D. 
\end{equation*}
For any $0 < \alpha < 1$, there exist two positive constants  $r_0(\alpha) < R_0(\alpha) $ depending only on $R$ such that if $r  <  r_0(\alpha)$,
then,  for some neighborhood $D_1$ of $\{z \in D; \; |z| = R_0(\alpha) \}$ and  for some positive constant $C$,  both independent of $v$ and $f$, 
\begin{equation*}
\| v\|_{H^1(D_1)} \le C  \big(\| v \|_{\bH(\Gamma)} + \| f\|_{L^2(\widetilde D)} \big)^{\alpha} \big( \| v \|_{H^1(\widetilde D)} + \| f\|_{L^2(\widetilde D)} \big)^{1 - \alpha} .  
\end{equation*}
\end{corollary}

\noindent{\bf Proof.} Set $v= u \circ T^{-1}$ with $T(z) = z^{1/m}$ for some $m\in  \mN$. Then $v \in H^1(T(\widetilde D))$ satisfies the equation 
$$
\Delta v + \tilde \sigma v = \tilde f \mbox{ in } T (\widetilde D), 
$$
where, by Lemma~\ref{lem-TO}, 
$$
\tilde \sigma v = T_*\sigma v / c \mbox{ and } \tilde f = T_*f / c, 
$$ 
for some positive constant $c$  such that $T_*I = c I$ (such a constant $c$ exists since $T$ is a conformal map). 
Corollary~\ref{cor1} is now a consequence of Lemma~\ref{lem-3S-1} by taking $m$ large enough.  \proofend

\medskip 
A variant of Lemma~\ref{lem-3S-2} is

\begin{lemma}\label{lem-3S-3} Let $0 < R \le R_1$, $m \in \mN$,  and define $D = T \big(B_{R} \cap  B(R_1, R_1) \big)$  where  $T(z) = z^{1/m}$. Set $\hat D =  \big\{ z \in  D; \; \frac{1}{20}R^{1/m} < |z| < R^{1/m} \big\}$ and $\Gamma = \big\{ z \in \partial D; \; \frac{1}{10}R^{1/m} < |z| < \frac{9}{10}R^{1/m} \big\}$. Let   $\sigma$ be a real bounded function defined in $\hat D$, $f \in L^2(\hat D)$, and   $v \in H^1(\hat D)$ be such that 
\begin{equation*}
\Delta v +  \sigma v = f \mbox{ in } \hat D. 
\end{equation*}
 For any $0 < \alpha < 1$, there exists $m_0 \in \mN$ depending only on $R$ such that if $m > m_0$ then,   for some neighborhood $D_1$ of $\{z \in D; \; |z| = \frac{1}{2}R^{1/m} \}$ and  for some positive constant $C$ both  independent of $v$ and $f$, 
\begin{equation*}
\| v\|_{H^1(D_1)} \le C  \big(\| v \|_{\bH(\Gamma)} + \| f\|_{L^2(\hat D)} \big)^{\alpha} \big( \| v \|_{H^1(\hat D)} + \| f\|_{L^2(\hat D)} \big)^{1 - \alpha}. 
\end{equation*}
\end{lemma}

\noindent{\bf Proof.} The proof of Lemma~\ref{lem-3S-3} is similar to the one of Lemma~\ref{lem-3S-2}. The details are left to the reader. \proofend 

\medskip 

As a consequence of Lemma~\ref{lem-3S-2}, we obtain 
\begin{corollary}\label{cor2} 
Let $0 < r <  R \le R_1$ and define $D = T\big(B_R \cap  B(R_1, R_1) \big)$  where  $T(z) = z^{1/m}$. Set $\widetilde D =  \big\{ z \in  D; \;  r < |z| < R \big\}$ and $\Gamma = \big\{ z \in \partial D; \; \frac{1}{10} R^{1/m}  < |z| < \frac{9}{10} R^{1/m} \big\}$. Let  $f \in L^2(\widetilde D)$,  $\sigma$ be a real bounded function defined in $\widetilde D$, and   $v \in H^1(\widetilde D)$ be such that 
\begin{equation*}
\Delta v +  \sigma v = f \mbox{ in } \widetilde D. 
\end{equation*}
 For any $0 < \alpha < 1$, there exist two positive constants $ r_0 (\alpha ) <  R_0(\alpha) $ depending only on $R$ such that if $r < r_0(\alpha)$ then,   for some neighborhood $D_1$ of $\{z \in D; \; |z| = R_0(\alpha) \}$ and  for some positive constant $C$ both  independent of $v$ and $f$, 
\begin{equation*}
\| v\|_{H^1(D_1)} \le C  \big(\| v \|_{\bH(\Gamma)} + \| f\|_{L^2(\widetilde D)} \big)^{\alpha} \big( \| v \|_{H^1(\widetilde D)} + \| f\|_{L^2(\widetilde D)} \big)^{1 - \alpha}. 
\end{equation*}
\end{corollary}

\noindent{\bf Proof.} The proof of Corollary~\ref{cor2} is similar to the one of Corollary~\ref{cor1}. The details are left to the reader. \proofend 

\medskip 
We are ready to  present the

\medskip  
\noindent{\bf Proof of Theorem~\ref{thm-CS-1-k}.} The proof of Theorem~\ref{thm-CS-1-k} is in the same spirit of the one of Theorem~\ref{thm-main-g}.  Applying Lemma~\ref{lem1}, we have 
\begin{equation}\label{part0-thm-CS-1-k}
\| u_{\delta} \|_{H^1(B_R)}\le C \Big(\delta^{-1/2} \|f \|_{L^2}^{1/2} \| u_\delta \|_{L^2(\supp f)}^{1/2} + \| f\|_{L^2}\Big). 
\end{equation} 
Here and in what follows in this proof, $C$ denotes a positive constant independent of $\delta$ and $f$. 

Let  $F$ and $G$ be the Kelvin transform with respect to $\partial B_{r_2}$ and $\partial B_{r_3}$. Define 
\begin{equation*}
u_{1, \delta} = u_\delta \circ F^{-1} \mbox{ in } \mR^2 \setminus B_{r_2} \quad \mbox{ and } \quad  
u_{2, \delta} = u_{1, \delta} \circ G^{-1} \mbox{ in } B_{r_3}. 
\end{equation*}
Set 
$$
S = B_{r_3} \setminus B_{r_2} \cap \big(B(x_2, r_0) \cup G \circ F (B(x_1, r_0) \cap B_{r_1}) \big).  
$$
By Lemma~\ref{lem-TO}, we have 
\begin{equation}\label{transmission1-pro2}
u_{1, \delta} = u_\delta \quad \mbox{ and }  \quad (1 + i \delta) \partial_\nu u_{1, \delta} = \partial_\nu u_{\delta} \mbox{ on } \partial B_{r_2} \setminus \partial S,  
\end{equation}
and 
\begin{equation}\label{transmission2-pro2}
u_{2, \delta} = u_{1, \delta} \quad  \mbox{ and }  \quad  \partial_\nu u_{2, \delta} = (1 + i \delta)\partial_\nu u_{1, \delta} \mbox{ on } \partial  B_{r_3} \setminus \partial S. 
\end{equation}
Combining \eqref{part0-thm-CS-1-k}, \eqref{transmission1-pro2}, and \eqref{transmission2-pro2} yields 
\begin{equation}\label{transmission1-pro2-1}
\|u_{1, \delta} - u_{\delta}\|_{\bH(\partial B_{r_2} \setminus \partial S)} \le C \delta \| u_\delta\|_{H^1(B_{r_3})}
\end{equation}
and  
\begin{equation}\label{transmission2-pro2-1}
\| u_{2, \delta} - u_{1, \delta}\|_{\bH( \partial B_{r_3} \setminus \partial S)} \le C \delta \| u_\delta \|_{H^1(B_{r_3})}.  
\end{equation}
Applying Lemma~\ref{lem-TO}, we have
$$
\dive\big((1 + i \delta) \nabla u_{1, \delta} \big) + k^2 u_{1, \delta}= 0 \mbox{ in } B_{r_3} \setminus B_{r_2}.  
$$
and 
$$
\Delta u_{2, \delta}  + k^2 u_{2, \delta}= 0 \mbox{ in } B_{r_3}  \setminus F \circ G (B(x_1, r_0) \cap B_{r_1}).  
$$
We derive that  
\begin{equation}\label{haha111}
\Delta u_{1, \delta} + k^2 u_{1, \delta}= \Big(1 - \frac{1}{1 + i \delta} \Big) k^2  u_{1, \delta} = - \frac{i \delta}{1 + i \delta } k^2  u_{1, \delta} \mbox{ in } B_{r_3} \setminus B_{r_2}. 
\end{equation}
Recall that 
\begin{equation}\label{haha121}
\Delta u_{\delta} + k^2  u_{\delta}= 0 \mbox{ in } (B_{r_3} \setminus B_{r_2} ) \setminus {\cal C}. 
\end{equation}
Denote $x_3 \in \partial B_{r_3}$  the image of $x_1$ by $F$ and set $\alpha = \gamma + 1/2$. We claim that   there exist three positive constants $r_0(\alpha), R_2(\alpha), R_3(\alpha)$, depending only on $r_2$ and $r_3$,  with 
$r_0(\alpha) <  R_2(\alpha) = R_2$, $r_0(\alpha) <  R_3(\alpha) = R_3$,  
such that, if $r_0  <r_0(\alpha)$,  then for some neighborhood $
D_{2}$ of $\Big\{z \in \mR^2 \setminus B_{r_2}; \;  |z - x_2| =R_2  \Big\}$,  and some neighborhood $
D_{3}$ of $ \Big\{z \in B_{r_3}; \;   |z - x_3| = R_3 \Big\}$
\begin{equation}\label{transmission1-pro2-2}
\| u_{1,\delta} - u_{\delta}\|_{H^1(D_{2})} \le C \delta^{\alpha} \| u_{\delta}\|_{H^1(B_{r_3})}
\end{equation}
and
\begin{equation}\label{transmission2-pro2-2}
\| u_{2,\delta} - u_{1, \delta}\|_{H^1(D_{3})} \le C \delta^{\alpha} \| u_{\delta}\|_{H^1(B_{r_3})}.  
\end{equation}
We first derive \eqref{transmission1-pro2-2} from Corollary~\ref{cor1}. 
For simplicity of notations, assume that $x_2 = (0, |x_2|)$. Applying Corollary~\ref{cor1} for $u_{1, \delta}(\cdot - x_2) - u_{\delta} (\cdot - x_2)$ , $R_1 = R = r_2$, we obtain assertion~\eqref{transmission1-pro2-2}. Assertion~\eqref{transmission2-pro2-2} can be obtained similarly by using Corollary~\ref{cor2}. 

Set 
$$
O_{2} = B_{r_2} \cup \{|z - x_2| <  R_2 \}, \quad O_{3} = B_{r_3} \setminus \{|z - x_3| <  R_3 \}, \quad \mbox{ and } \quad O = O_3 \setminus O_2. 
$$
It follows from \eqref{transmission1-pro2-1} and \eqref{transmission1-pro2-2}  that 
\begin{equation} \label{claim1-pro2}
\| u_{1,\delta} - u_{\delta}\|_{\bH(\partial O_{2})} \le C \delta^{\alpha} \| u_{\delta}\|_{H^1(B_{r_3})}
\end{equation}
and from \eqref{transmission2-pro2-1} and \eqref{transmission2-pro2-2} that
\begin{equation} \label{claim2-pro2}
\| u_{2,\delta} - u_{1, \delta}\|_{\bH(\partial O_{3})} \le C \delta^{\alpha} \| u_{\delta}\|_{H^1(B_{r_3})}.  
\end{equation}
Define 
\begin{equation}\label{def-Udelta-1}
{\cal U}_\delta  = \left\{\begin{array}{cl} u_\delta & \mbox{ in } \mR^d \setminus O_3, \\[6pt]
u_{2, \delta} -   (u_{1, \delta} - u_{\delta}) & \mbox{ in } O, \\[6pt]
u_{2, \delta} &  \mbox{ in } O_2. 
\end{array}\right. 
\end{equation}
Then ${\cal U}_\delta \in H^1 \big(B_R \setminus \partial O \big)$ for all $R>0$ and $U_\delta$ is an outgoing solution of the equation
\begin{equation*}
\dive (\hat A \nabla {\cal U}_\delta) + k^2 \hat \Sigma  {\cal U}_{\delta} = f \mbox{ in } \mR^2 \setminus  \partial O, 
\end{equation*}
by the definition of $(\hat A, \hat \Sigma)$ and Lemma~\ref{lem-TO}. 
This implies
\begin{equation}\label{part0-pro2}
\|{\cal U}_\delta \|_{H^1(B_R \setminus \partial O)} \\[6pt] 
\le C \Big( \|[{\cal U}_\delta] \|_{H^{1/2}(\partial O)} + \|[\hat A \nabla  {\cal U}_\delta  \cdot \nu] \|_{H^{-1/2}(\partial O)} + \| f\|_{L^2} \Big). 
\end{equation}
From the definition of ${\cal U}_\delta$ in \eqref{def-Udelta-1}, we derive from   \eqref{claim1-pro2} and \eqref{claim2-pro2} that 
\begin{equation}\label{transmission3-pro2}
\|[{\cal U}_\delta] \|_{H^{1/2}(\partial O)} + \|[\hat A \nabla {\cal U}_\delta  \cdot \nu] \|_{H^{-1/2}(\partial O)} \le  C \delta^{\alpha} \| u_\delta \|_{H^1(B_{r_3})}. 
\end{equation}
Combining \eqref{part0-pro2} and  \eqref{transmission3-pro2} and using \eqref{part0-thm-CS-1-k} yield, for large $R> 0$ such that $\supp f \subset B_R$, 
\begin{equation*}
\| {\cal U}_\delta\|_{H^1(B_{R} \setminus \partial O)} \le  C_R \delta^{\alpha} \Big( \delta^{-1/2}\| {\cal U}_\delta \|_{L^2(B_R \setminus B_{r_3})}^{1/2} \|f \|_{L^2}^{1/2} + \| f\|_{L^2}\Big) + C \|f \|_{L^2}; 
\end{equation*}
which implies, since $\alpha > 1/2$,  
\begin{equation}\label{t1}
\| {\cal U}_\delta\|_{H^1(B_{R} \setminus \partial O)} \le  C_R  \|f \|_{L^2}. 
\end{equation}
By \eqref{part0-thm-CS-1-k}, we obtain 
\begin{equation}\label{ud-improve2}
\| u_\delta\|_{H^1(B_{R} )} \le C_R \delta^{-1/2}\| f\|_{L^2}. 
\end{equation}
We derive from \eqref{transmission3-pro2} that
\begin{equation}\label{t2}
\|[{\cal U}_\delta] \|_{H^{1/2}(\partial O)} + \|[\hat A \nabla {\cal U}_\delta  \cdot \nu] \|_{H^{-1/2}(\partial O)} \le  C \delta^{\alpha - 1/2} \| f\|_{L^2}. 
\end{equation}
From \eqref{t1} and \eqref{t2}, without loss of generality,  one may assume that ${\cal U}_{\delta} \to {\cal U}$ weakly in $H^1\big(B_R \setminus \partial O \big)$  as $\delta \to 0$ for any $R>0$; moreover, ${\cal U} \in H^1_{\loc}(\mR^d)$  is the unique outgoing solution to the equation  
\begin{equation*}
\dive (\hat A \nabla  {\cal U}) + k^2 \hat \Sigma  {\cal U} = f \mbox{ in } \mR^2. 
\end{equation*}
Hence ${\cal U} = \cU$ in $\mR^2$. Since the limit is unique,  the convergence holds for the whole family $({\cal U}_{\delta})$ as $\delta \to 0$. 

\medskip 
By considering ${\cal U}_\delta - \cU$, one obtains the rate of the convergence as in the proof of Theorem~\ref{thm-main-g}. The proof is complete. \proofend

\begin{remark}\label{rem2-power} The power of $u_\delta$ defined by $\dsp \int_{\Omega_2 \setminus \Omega_1} \delta |\nabla u_\delta|^2$
is finite in the setting considered in Theorem~\ref{thm-CS-1} by \eqref{ud-improve2}. 
\end{remark}

\medskip  
\noindent{\bf Proof of Theorem~\ref{thm-CS-1}.} The proof of Theorem~\ref{thm-CS-1} is similar to the one of Theorem~\ref{thm-CS-1-k} even simpler. The details are left to the reader. \proofend

\subsection{A remark on superlensing using complementary media}\label{sect-CM-Superlensing}

In this section, we revisit  the construction of lenses  using complementary proposed in \cite{Ng-Superlensing} and \cite{Ng-Negative-Cloaking} which has roots from \cite{NicoroviciMcPhedranMilton94, PendryCylindricalLenses, PendryRamakrishna}. We first consider the   quasistatic regime. To magnify $M$-times the region $B_{\tau_0}$,  for some $\tau_0> 0$ and $M> 1$, following \cite{NicoroviciMcPhedranMilton94, PendryCylindricalLenses, PendryRamakrishna} one puts a lens in $B_{r_2} \setminus B_{\tau_0}$ whose medium is  characterized by matrix $-I$  with $r_2^2/ \tau_0^2 = M$. The lens construction in \cite{Ng-Superlensing} is related to but different from this.
Our lens contains {\bf two parts}. The first one is given by 
\begin{equation}\label{SCM-first-part}
-I  \quad \mbox{ in } B_{r_2} \setminus B_{r_1}
\end{equation} 
and the second one is 
\begin{equation}\label{SCM-second-part}
I \quad \mbox{ in } B_{r_1} \setminus B_{\tau_0}. 
\end{equation}
Here $r_1$ and $r_2$ are such that 
\begin{equation}\label{SCM-def-r3}M \tau_0 = r_2 \quad \mbox{ and } \quad r_3/ r_1  = M \quad \mbox{ where } 
r_3 := r_2^2 / r_1. 
\end{equation}
Other choices for the first and the second layers are possible via the concept of complementary media (see \cite{Ng-Superlensing}). 
Given  $f \in L^2(\mR^2)$ with compact support such that $\int_{\mR^2} f = 0$, and $\supp f \cap B_{r_3}  = \O$,   we showed in \cite{Ng-Superlensing},  that 
\begin{equation*}
u_\delta \to {\cU} \mbox{ in } \mR^2 \setminus B_{r_3} \mbox{ as } \delta \to 0 
\end{equation*} 
Here $u_\delta$ and $\cU$ are the unique solutions in $W^1(\mR^2)$ of the equations 
$$
\dive(s_\delta A \nabla u_\delta) = f \mbox{ in } \mR^2 \quad \mbox{ and } \quad \Delta u = f \mbox{ in } \mR^2, 
$$
where 
\begin{equation*}
A = \left\{\begin{array}{cl} 
a(x/ M) &  \mbox{ in } B_{M r_1}, \\[6pt]
I & \mbox{ otherwise}. 
\end{array}\right. \quad \mbox{ and } \quad 
s_\delta = \left\{\begin{array}{cl}  -1 -  i \delta &  \mbox{ in } B_{r_2} \setminus B_{r_1}, \\[6pt]
1 & \mbox{ otherwise}.  
\end{array}\right. 
\end{equation*}
for arbitrary uniformly elliptic matrix-valued function $a$ defined in $B_{r_1}$: superlensing is achieved.  
In fact, in \cite{Ng-Superlensing}, we considered bounded setting, nevertheless, the results stated here holds with similar proof. 
In comparison with the schemes suggested previously, the second part is added.  
The second layer  can be chosen thinner (see \cite{Ng-Superlensing}). Nevertheless, 
as a consequence of Theorem~\ref{thm-CS-1}, 
the second layer in \eqref{SCM-second-part} is necessary. Indeed, we have

\begin{proposition}\label{pro-lensing}
Let $0 < r_0 <  r_1 < r_2 < r_3$,  $x_1 \in \partial B_{r_1}$, and $x_2  \in \partial B_{r_2}$ and let $a_c$ be a symmetric uniformly elliptic matrix-valued function defined in $B(x_1, r_0) \cap B_{r_1}$. 
Define 
\begin{equation}\label{def-AS-CS*}
A_c = \left\{\begin{array}{cl} 
a_c &  \mbox{ in } B(x_1, r_0) \cap B_{r_1}, \\[6pt]
I & \mbox{ otherwise}. 
\end{array}\right. \quad 
\mbox{ and } \quad 
s_\delta = \left\{\begin{array}{cl}  -1 -  i \delta &  \mbox{ in } B_{r_2} \setminus B_{r_1}, \\[6pt]
1 & \mbox{ otherwise}.  
\end{array}\right. 
\end{equation} 
Given $f \in L^2(\mR^2)$ with compact support such that $\int_{\mR^2} f = 0$ and $\supp f \cap B_{r_3} = \O$ with $r_3 = r_2^2/ r_1$, let $u_\delta, \cU \in W^1(\mR^2)$ be the unique solutions of \eqref{def-udelta-CS} and \eqref{def-U-CS} respectively.  There exists $r_* > 0$ depending only  only on $r_1$ and $r_2$,  such that if $r_0 < r_*$ then 
\begin{equation*}
u_\delta  \to \cU \mbox{ in } L^2(\Omega \setminus B_{r_3}).
\end{equation*}
\end{proposition}

For an observer outside $B_{r_3}$, the medium $s_0A_c$ in $B_{r_3}$ looks like $I$ in $B_{r_3}$: the object $a_c$ is disappeared.  Thus the superlensing device becomes a cloak for small objects close to it. Similar conclusion also holds for the finite frequency regime as a consequence of Theorem~\ref{thm-CS-1-k}. 

\begin{remark} Using Theorem~\ref{thm-main-g}, in the same spirit with what presented in this section, one could argue to show that the necessity of the second layer in the lens construction suggested in \cite{Ng-Superlensing} in three dimensions.   
\end{remark}

\subsection{A remark on cloaking using complementary media} \label{sect-CM-Cloaking}

This section deals with cloaking using complementary media. We show that a modification proposed in \cite{Ng-Negative-Cloaking} (see also \cite{MinhLoc2}) from the suggestion   in \cite{LaiChenZhangChanComplementary} on cloaking using complementary media is necessary: without a modification, cloaking might be not achieved. We only consider here the two dimensional finite frequency regime. The two dimensional quasistatic case holds similarly.  

\medskip 
We first describe how to cloak  the region $B_{2r_2} \setminus B_{r_2}$ for some $r_{2}> 0$ using complementary media as proposed in \cite{Ng-Negative-Cloaking}. Assume that the medium in  $B_{2r_2} \setminus B_{r_2}$ is characterized by a symmetric uniformly elliptic matrix-valued function  $a_c$ and a real function $\sigma_c$ bounded above and below by positive constants. We only consider here the two dimensional case; even the construction in three dimensions hold similarly.  The idea suggested by Lai et al. in \cite{LaiChenZhangChanComplementary} is to construct a complementary media in $B_{r_2} \setminus B_{r_1}$ for some $0 < r_1< r_2$. Our cloak proposed in  \cite{Ng-Negative-Cloaking} is related to but different from \cite{LaiChenZhangChanComplementary}. It  consists of {\bf two parts}. The first one, in $B_{r_2} \setminus B_{r_1}$,  makes use of reflecting complementary media to cancel the effect of the cloaked region and the second  one in $B_{r_1}$  is to fill the space which ``disappears" from the cancellation by the homogeneous medium $(I, 1)$. 
For the first part, we also modified the strategy in \cite{LaiChenZhangChanComplementary}. Instead of $B_{2r_2} \setminus B_{r_2}$, we consider $B_{r_3} \setminus B_{r_2}$ for some $r_3 > 0$ as the cloaked region in which the medium is given by 
\begin{equation}\label{def-b}
(a_1, \sigma_1) = \left\{ \begin{array}{cl} (a_c, \sigma_c) & \mbox{ in } B_{2 r_2} \setminus B_{r_2}, \\[6pt]
(I, 1) & \mbox{ in } B_{r_3} \setminus B_{2 r_2}. 
\end{array} \right. 
\end{equation} 
The complementary medium in $B_{r_2} \setminus B_{r_1}$ with $r_1 = r_2^2/ r_3$ is
\begin{equation*}
\big(-{F^{-1}}_*a_1, -{F^{-1}}_*\sigma_1 \big), 
\end{equation*}
where $F: B_{r_2} \setminus \bar B_{r_1} \to B_{r_3} \setminus \bar B_{r_2}$ is the Kelvin transform with respect to $\partial B_{r_2}$.  Concerning the second part, the medium in $B_{r_1}$ is given by
\begin{equation}\label{CCM-choice-A}
\big(  I, r_3^2/ r_1^2 \big). 
\end{equation}
Set
\begin{equation}\label{def-AS-1}
(A, \Sigma) = \left\{\begin{array}{cl} 
(a_1, \sigma_1) &  \mbox{ in } B_{r_3} \setminus B_{r_2}, \\[6pt]
\big({F^{-1}}_*a_1,  {F^{-1}}_*\sigma_1 \big)& \mbox{ in } B_{r_2} \setminus B_{r_1}, \\[6pt]
\Big( I,  r_3^2/r_1^{2} \Big)& \mbox{ in } B_{r_1}, \\[6pt]
(I, 1) & \mbox{ otherwise}, 
\end{array}\right. \quad \mbox{ and } \quad 
s_\delta = \left\{\begin{array}{cl}  -1 -  i \delta &  \mbox{ in } B_{r_2} \setminus B_{r_1}, \\[6pt]
1 & \mbox{ otherwise}. 
\end{array}\right. 
\end{equation}
Given  $f \in L^2(\mR^2)$ with compact support such that $\supp f \cap B_{r_3}  = \O$,   we showed in \cite[Theorem 1]{MinhLoc2} (see \cite[Theorem 1]{Ng-Negative-Cloaking} for the case $k=0$) that, if $r_3/r_2$ is large enough, 
\begin{equation*}
u_\delta \to {\cU} \mbox{ in } \mR^2 \setminus B_{r_3} \mbox{ as } \delta \to 0. 
\end{equation*} 
Here $u_\delta$ and $\cU$ are the unique outgoing solutions in $H^1_{\loc}(\mR^2)$ of 
$$
\dive(s_\delta A \nabla u_\delta) + s_0 k^2 \Sigma u_\delta = f \mbox{ in } \mR^2  \quad \mbox{ and } \quad \Delta \cU + k^2 \cU = f \mbox{ in } \mR^2.
$$
For an observer outside $B_{r_3}$,  the medium $(s_0A, s_0\Sigma)$ acts like $(I,1)$: cloaking is achieved.  

\medskip 
The following proposition show that  it is necessary to  extend $(a_c, \sigma_c)$ by $(I, 1)$ in $B_{r_3} \setminus B_{2 r_2}$ in the construction of cloaking device given above.  

\begin{proposition}\label{pro-cloaking} Let $0 < r_0 <  r_1 < r_2 $,  and $x_3 \in \partial B_{r_3}$ with $r_3 = r_2^2/ r_1$. Let $a_c$ be a symmetric uniformly elliptic matrix-valued function and $\sigma_c$ be a positive function bounded below by a positive constant both defined in $B(x_3, r_0) \cap B_{r_3}$. Define $(A, \Sigma)$ by \eqref{def-AS-1} where 
\begin{equation*}
(a_1, \sigma_1) = \left\{\begin{array}{cl} (a_c, \sigma_c) &  \mbox{ in } B(x_3, r_0) \cap B_{r_3}, \\[6pt]
(I, 1) & \mbox{ in } (B_{r_3} \setminus B_{r_2}) \setminus B(x_3, r_0).  
\end{array}\right.
\end{equation*}
Given $f \in L^2(\mR^2)$ with compact support  and $\supp f \cap B_{r_3} = \O$,  let $u_\delta \in H^1_{\loc}(\mR^2)$ be the unique outgoing solution to the equation 
\begin{equation*}
\dive(s_\delta A \nabla u_\delta )  + s_0 k^2 \Sigma u_\delta  = f \mbox{ in } \mR^2,  
\end{equation*}
where $(A, \Sigma)$ and $s_\delta$ are given by \eqref{def-AS-1}. 
There exists $r_* > 0$ depending only on $r_1$ and $r_2$ such that if $r_0 < r_*$, then 
\begin{equation}\label{tt1}
u_\delta  \to \cU \mbox{ in } L^2_{\loc}(\mR^2 \setminus B_{r_3}). 
\end{equation}
Here $\cU \in H^1_{\loc}(\mR^2)$ is the unique  outgoing solution to the equation  
\begin{equation}\label{tt2}
\dive(\cA \nabla \cU)  + k^2 \cS \cU   = f \mbox{ in } \Omega \mbox{ where } (\cA, \cS) = \left\{\begin{array}{cl} (a_c, \sigma_c) & \mbox{ in } B(x_0, r_0) \cap B_{r_3}, \\[6pt]
(I, 1) & \mbox{ otherwise}. 
\end{array}\right.
\end{equation}
\end{proposition}

As a consequence of Proposition~\ref{pro-cloaking}, the object $(a_c, \sigma_c)$ in $B(x_3, r_0) \cap B_{r_3}$ is not cancelled by its complementary media by \eqref{tt1} and \eqref{tt2}: cloaking is {\bf not} achieved.

\medskip 

\noindent{\bf Proof.} The proof is similar to the one of Theorem~\ref{thm-CS-1-k} even simpler. We only sketch the proof. We have, by Lemma~\ref{lem1}, 
\begin{equation*}
\| u_\delta \|_{H^1(B_R)}\le C_R \Big(\delta^{-1/2} \|f \|_{L^2}^{1/2} \| u_\delta \|_{L^2(\supp f)}^{1/2} + \| f\|_{L^2}\Big). 
\end{equation*} 
Here and in what follows in this proof, $C$ denotes a positive constant independent of $\delta$ and $f$. 
Define 
\begin{equation*}
u_{1, \delta} = u_\delta \circ F^{-1} \mbox{ in } \mR^2\setminus B_{r_2} \quad \mbox{ and } \quad 
u_{2, \delta} = u_{1, \delta} \circ G^{-1} \mbox{ in } B_{r_3}, 
\end{equation*}
where $F$ and $G$ are the Kelvin transforms with respect to $\partial B_{r_2}$ and $\partial B_{r_3}$ respectively. By Lemma~\ref{lem-TO}, as in \eqref{transmission1-pro2-1} and  \eqref{transmission2-pro2-1} we have 
\begin{equation}\label{transmission1-pro2-1-*}
\|u_{1, \delta} - u_{\delta}\|_{\bH(\partial B_{r_2})} \le C \delta \| u_\delta\|_{H^1(B_{r_3})}
\end{equation}
and  
\begin{equation}\label{transmission2-pro2-1-*}
\| u_{2, \delta} - u_{1, \delta}\|_{\bH( \partial B_{r_3} \setminus B(x_0, r_0))} \le C \delta \| u_\delta \|_{H^1(B_{r_3})}.  
\end{equation}
Applying Lemma~\ref{lem-TO}, we have
$$
(1 + i \delta) \Delta u_{1, \delta}  + k^2 u_{1, \delta}= 0 \mbox{ in } (B_{r_3} \setminus B_{r_2}) \setminus B(x_3, r_0)
$$
and
$$
\Delta u_{2, \delta} + k^2  u_{2, \delta}= 0 \mbox{ in } B_{r_3}. 
$$
We derive that  
\begin{equation}\label{haha1111}
\Delta u_{1, \delta} + k^2 u_{1, \delta}= \Big(1 - \frac{1}{1 + i \delta} \Big) k^2  u_{1, \delta} = - \frac{i \delta}{1 + i \delta } k^2  u_{1, \delta} \mbox{ in } (B_{r_3} \setminus B_{r_2}) \setminus B(x_3, r_0). 
\end{equation}
For simplicity of notations, we assume that $x_3 = (0, - |x_3|)$.
Applying Corollary~\ref{cor2} for $u_{1, \delta}(\cdot - x_3) - u_{2, \delta}(\cdot - x_3)$, $R = r_2/2 < R_1 = r_3$, and  $\alpha = 3/4$, there exist two positive constants $ r_* < R_0$ such that if $r_0 < r_*$ then for some neighborhood $D$ of $B_{r_3} \cap \big\{z; |z - x_3| = R_0  \big\}$
$$
\| u_{1, \delta} - u_{2, \delta} \|_{H^1(D)} \le C \delta^\alpha \| u_\delta\|_{H^1(B_{r_3})}. 
$$
Set $O = (B_{r_3} \setminus B_{r_2} )  \setminus \big\{z \; ;  |z - x_3| < R_0 \big\}$ and define 
\begin{equation*}
{\cal U}_\delta  = \left\{\begin{array}{cl} 
u_{2, \delta} + u_{\delta} -   u_{1, \delta}  & \mbox{ in } O, \\[6pt]
u_{2, \delta} &  \mbox{ in } B_{r_2}, \\[6pt]
u_\delta & \mbox{ otherwise.}
\end{array}\right. 
\end{equation*}
Then ${\cal U}_\delta \in H^1 \big(B_R \setminus \partial O  \big)$ for all $R> 0$ and $U_\delta$ is an outgoing solution of the equation
\begin{equation*}
\dive (\hat A \nabla {\cal U}_\delta) + k^2 \hat \Sigma  {\cal U}_{\delta} = f \mbox{ in } \mR^2 \setminus \partial O.  
\end{equation*}
and,  as in \eqref{transmission3-pro2}, 
\begin{equation*}
\|[{\cal U}_\delta] \|_{H^{1/2}(\partial O)} + \|[\hat A \nabla {\cal U}_\delta  \cdot \nu] \|_{H^{-1/2}(\partial O)} \le  C \delta^{\alpha} \| u_\delta \|_{H^1(B_{r_3})}. 
\end{equation*}
It follows that, for large $R> 0$ such that $\supp f \subset B_R$, 
\begin{equation*}
\| {\cal U}_\delta\|_{H^1(B_{R} \setminus \partial O )} \le  C_R \delta^{\alpha} \Big( \delta^{-1/2}\| {\cal U}_\delta \|_{L^2(B_R \setminus B_{r_3})}^{1/2} \|f \|_{L^2(\Omega)}^{1/2} + \| f\|_{L^2}\Big) + C \|f \|_{L^2}.
\end{equation*}
Since $\alpha = 3/4 > 1/2$,  as in the proof of  Theorem~\ref{thm-CS-1-k}, one deduces that $U_\delta$ is bounded in $H^1(B_R \setminus \partial O)$, 
\begin{equation*}
\|[{\cal U}_\delta] \|_{H^{1/2}(\partial O)} + \|[\hat A \nabla {\cal U}_\delta  \cdot \nu] \|_{H^{-1/2}(\partial O)} \to 0, 
\end{equation*}
 and $U_\delta \to {\cal U}$ in $H^1_{\loc} (\mR^2 \setminus B_{r_3})$ as $\delta \to 0$.   The details are left to the reader.  \proofend

\begin{remark} It would be interesting to understand the cooperation and the combat of various cloaking devices using NIMs putting together as raised in \cite{Ross}. 

\end{remark}

\providecommand{\bysame}{\leavevmode\hbox to3em{\hrulefill}\thinspace}
\providecommand{\MR}{\relax\ifhmode\unskip\space\fi MR }
\providecommand{\MRhref}[2]{%
  \href{http://www.ams.org/mathscinet-getitem?mr=#1}{#2}
}
\providecommand{\href}[2]{#2}

\end{document}